\documentclass[paper]{JHEP3}
\pdfoutput=1
\usepackage{amsmath,amssymb,amsthm,amscd,graphicx}
\input epsf.sty

\theoremstyle{definition}

\newcommand{\CC}{{\cal C}}

\newcommand{\CF}{{\cal F}}

\newcommand{\CI}{{\cal I}}

\newcommand{\CN}{{\cal N}}
\newcommand{\CO}{{\cal O}}
\newcommand{\CP}{{\cal P}}

\newcommand{\CR}{{\cal R}}

\newcommand{\CW}{{\cal W}}

\def\IZ{{\mathbb Z}}
\def\IR{{\mathbb R}}
\def\IC{{\mathbb C}}
\def\IP{{\mathbb P}}

\def\IS{{\mathbb S}}
\def\IF{{\mathbb F}}

\newcommand{\tr}{{\rm Tr}}
\newcommand{\re}{{\rm e}}
\newcommand{\ri}{{\rm i}}
\newcommand{\rd}{{\rm d}}
\renewcommand{\d}{\partial}

\newcommand{\be}{\begin{equation}}
\newcommand{\ee}{\end{equation}}
\newcommand{\ba}{\begin{aligned}}
\newcommand{\ea}{\end{aligned}}
\newcommand{\ben}{\begin{eqnarray}\displaystyle}
\newcommand{\een}{\end{eqnarray}}

\newcommand{\p}{\partial}
\newcommand{\sectiono}[1]{\section{#1}\setcounter{equation}{0}}

\newdimen\tableauside\tableauside=1.0ex
\newdimen\tableaurule\tableaurule=0.4pt
\newdimen\tableaustep
\def\phantomhrule#1{\hbox{\vbox to0pt{\hrule height\tableaurule width#1\vss}}}
\def\phantomvrule#1{\vbox{\hbox to0pt{\vrule width\tableaurule height#1\hss}}}
\def\sqr{\vbox{%
  \phantomhrule\tableaustep
  \hbox{\phantomvrule\tableaustep\kern\tableaustep\phantomvrule\tableaustep}%
  \hbox{\vbox{\phantomhrule\tableauside}\kern-\tableaurule}}}
\def\squares#1{\hbox{\count0=#1\noindent\loop\sqr
  \advance\count0 by-1 \ifnum\count0>0\repeat}}
\def\tableau#1{\vcenter{\offinterlineskip
  \tableaustep=\tableauside\advance\tableaustep by-\tableaurule
  \kern\normallineskip\hbox
    {\kern\normallineskip\vbox
      {\gettableau#1 0 }%
     \kern\normallineskip\kern\tableaurule}%
  \kern\normallineskip\kern\tableaurule}}
\def\gettableau#1{\ifnum#1=0\let\next=\null\else
\squares{#1}\let\next=\gettableau\fi\next}

\newcommand{\figref}[1]{Fig.~\protect\ref{#1}}

\preprint{HU-EP-10/39}

\title{From weak to strong coupling in ABJM theory}

\author{
Nadav Drukker$^a$, Marcos Mari\~no$^{b,c}$ and Pavel Putrov$^c$
\\
$^a$Institut f\"ur Physik, Humboldt-Universit\"at zu Berlin,\\
\phantom{$^a$}%
Newtonstra\ss e 15, D-12489 Berlin, Germany\\
$^b$D\'epartement de Physique Th\'eorique et $^c$Section de Math\'ematiques,\\
\phantom{$^b$}%
Universit\'e de Gen\`eve, Gen\`eve, CH-1211 Switzerland\\
\\
\email{drukker@physik.hu-berlin.de}, \quad
\email{marcos.marino@unige.ch}, \quad
\email{pavel.putrov@unige.ch}
}

\abstract{The partition function of $\CN=6$ supersymmetric Chern--Simons-matter theory 
(known as ABJM theory) on $\IS^3$, as well as certain Wilson loop observables, are captured by a zero dimensional super-matrix 
model. This super--matrix model is closely related to a matrix model describing topological Chern--Simons theory on a lens 
space. We explore further these recent observations and extract more exact results in ABJM theory from the matrix model. 
In particular we calculate the planar free energy, which matches at strong coupling the classical 
IIA supergravity action on AdS$_4\times\IC\IP^3$ and gives the correct $N^{3/2}$ scaling 
for the number of degrees of freedom of the M2 brane theory. Furthermore we find contributions 
coming from world-sheet instanton corrections in $\IC\IP^3$. We also calculate non-planar corrections, both 
to the free energy and to the Wilson loop expectation values. This matrix model 
appears also in the study of topological strings on a toric Calabi--Yau manifold, and an intriguing 
connection arises between the space of couplings of the planar ABJM theory and the 
moduli space of this Calabi--Yau. In particular it suggests that, in addition to the 
usual perturbative and strong coupling (AdS) expansions, a third natural expansion 
locus is the line where one of the two 't~Hooft couplings vanishes and the 
other is finite. This is the conifold locus of the Calabi--Yau, and leads to an expansion around 
topological Chern--Simons theory. We present some explicit results for the 
partition function and Wilson loop observables around this locus.}

\begin{document}

\newpage
\sectiono{Introduction and summary}
\label{sec:intro}

The discovery of Aharony, Bergman, Jafferis and Maldacena (ABJM) 
of the world-volume theory of coincident M2-branes \cite{abjm}
(following Bagger-Lambert and Gustavsson \cite{gus,bl}) provides a new interacting 
field theory with well defined weak and strong coupling expansions. A great deal of 
effort has been given to studying these two limits of the theory: three dimensional 
$\CN=6$ supersymmetric Chern--Simons-matter and type IIA string theory on 
AdS$_4\times\IC\IP^3$ (or M-theory on AdS$_4\times\IS^7/\IZ_k$). 
For better or worse, both descriptions of the theory are much harder than the 
D3-brane analog: 4d $\CN=4$ SYM and type IIB string theory on AdS$_5\times \IS^5$. 
At weak coupling perturbative calculations in ABJM theory are rather subtle and 
for many quantities are in even powers of the coupling, while at strong coupling the geometry of 
$\IC\IP^3$ is more complicated than $\IS^5$ and has, for example, non-trivial 
2-cycles.

An important breakthrough, which is the underpinning of the present study, 
was the work of Kapustin, Willett and Yaakov \cite{kapustin}, who use the localization techniques of \cite{pestun} to reduce the calculation 
of certain quantities in the gauge theory on $\IS^3$ to finite dimensional matrix integrals\footnote{Similar results apply also to other 3d theories with $\CN=2$ supersymmetry 
\cite{kapustin,kapustin2}.}.
These matrix integrals can be evaluated in a systematic expansion in 
$1/N$. Indeed, they have a natural supergroup structure, {\em i.e.}, they are 
super-matrix models \cite{dt,mp}, and are related to some previously studied 
bosonic matrix models \cite{mm,akmv} by analytical continuation \cite{mp}.

The solution of this matrix model allowed for the evaluation of the first exact 
interpolating function in this theory \cite{mp} giving a closed form expression for the 
expectation value of the $1/2$ BPS Wilson loop operator of \cite{dt} at all values of 
the coupling. This expression derived from the matrix-model reduction 
of the gauge theory reproduces exactly the known leading strong coupling 
result, the classical action of a macroscopic string in AdS$_4$.

The purpose of this paper is to explore further what can be learnt from the matrix model 
and its solution to the understanding of the physical 3d gauge theory and its string/M-theory 
dual.

This is a broad subject, connected through the matrix model to special geometry, 
Chern--Simons (CS) theory, topological strings and more. One of the avenues we 
explore is the relation between the moduli space of the matrix model and the 
space of couplings of the gauge theory. It is very useful to consider the generalization 
of the gauge theory where the rank of the two gauge groups are not equal 
\cite{abj}.%
\footnote{Though commonly known as ABJ theory, for simplicity we still call the theory 
with this extra parameter as ABJM theory. When specializing to the case of equal 
rank we refer to it as the ``ABJM slice''.}
The space of couplings is two dimensional and upon complexification,  it matches 
the moduli space of the Riemann surface solving the planar matrix model. This surface is also the mirror to a well studied 
toric Calabi--Yau manifold known as local $\IF_0$, where $\IF_0=\IP^1 \times \IP^1$ is a Hirzebruch surface. 
As we review in Section~\ref{sec:pf}, this moduli space 
has three special loci: the orbifold point, the large radius limit and the conifold locus.

These can be identified in the gauge theory respectively as the weakly coupled gauge 
theory, the strongly coupled theory described by string theory on AdS, and lastly the conifold locus 
is where the rank of one of the gauge group vanishes, so ABJM theory reduces 
to topological CS theory \cite{witten}. The first two are known duality frames with the AdS/CFT 
rules on how to evaluate observables on both sides. The simplicity of the conifold 
locus suggests that there should be another duality frame where ABJM theory 
is considered as a deformation of topological CS theory. We explore this in 
Section~\ref{sec:conifold}, where we calculate the partition function and Wilson loop 
observables around this point. It would be very interesting to learn how to 
calculate other quantities in this regime.

We present the matrix model for the ABJM theory and that for CS theory on 
the lens space $L(2,1)=\IS^3/\IZ_2$ in the next section. The matrix model of ABJM 
has an underlying $U(N_1|N_2)$ symmetry while that of the lens space has 
$U(N_1+N_2)$ symmetry, which in both cases are broken to 
$U(N_1)\times U(N_2)$. It is easy to see that the expressions 
for them are related by analytical continuation of $N_2\to-N_2$, or analogously 
a continuation of the 't Hooft coupling $N_2/k\to -N_2/k$ (which may be attributed 
to the negative level of the CS coupling of this group in the ABJM theory). 
We can then go on to study the lens space model and analytically continue to 
ABJM at the end.

Conveniently, the lens space matrix model has been studied in the past 
\cite{akmv,hy,mpp,mp}. The planar resolvent is known in closed form and the 
expressions for its periods are given as power series at special points in 
moduli space. We review the details of this matrix model and its solution in 
Sections~\ref{sec:MM} and~\ref{sec:pf}.

The matrix model of ABJM theory was derived by localization: it captures in a 
finite dimensional integral all observables of the full theory which preserve 
certain supercharges. At the time it was derived in \cite{kapustin}, the only 
such observable (apart for the vacuum) was the $1/6$ BPS Wilson loop 
constructed in \cite{dp,cw,rey} and $1/2$ BPS vortex loop operators 
\cite{vortex}. Indeed, the expectation value of the $1/6$ BPS Wilson loop 
can be expressed as an observable in the ABJM matrix model, and by 
analytical continuation in the lens space model.

Another class of Wilson loop operators, which preserve $1/2$ of the supercharges, 
was constructed in \cite{dt} and studied further in \cite{ll}. It is the dual of the 
most symmetric classical string solution in AdS$_4\times\IC\IP^3$. This Wilson 
loop is based on a super-connection in space-time and reduces upon localization 
to the trace of a supermatrix in the ABJM matrix model \cite{dt}. The different 
$1/2$ BPS Wilson loops are classified by arbitrary representations of 
the supergroup $U(N_1|N_2)$, and the $1/6$ BPS ones are classified 
by a pair of representations%
\footnote{Special combinations of representations of $U(N_1)\times U(N_2)$ 
are also representations of $U(N_1|N_2)$, and in this case the $1/6$ BPS 
and $1/2$ BPS loops will have the same expression in the matrix model and the 
same VEVs. The proof of localization for the $1/2$ BPS loop \cite{dt} relied 
on this equivalence.}
of $U(N_1)$ and $U(N_2)$ 
We will mostly concern ourselves with the $1/2$ BPS Wilson loop in the 
fundamental representation of $U(N_1|N_2)$ and the $1/6$ BPS Wilson 
loop in the fundamental representation of $U(N_1)$. The exception is 
Section~\ref{sec:giant} and Appendix~\ref{sec:cs-giant}, where we study the 
$1/2$ BPS Wilson loop in large symmetric and antisymmetric representations. 
There we also make contact with the vortex loop operators of \cite{vortex}.

Of course, the natural observables in CS theory are the partition function and Wilson 
loops, so these quantities were also studied earlier in the matrix models 
of CS (see, for example, \cite{mm,akmv,tierz,hy,leshouches,dtierz}). This information is encoded in different period integrals on the surface 
solving the matrix model, as we explain in Section~\ref{sec:wl-intro}. It turns out 
that the $1/6$ BPS loop is captured by a period integral around one of the two 
cuts in the planar solution and the $1/2$ BPS Wilson loop by a period 
integral around both cuts, or alternatively, around the point at infinity, and is 
much easier to calculate \cite{mp}.

With all this machinery presented in Sections~\ref{sec:MM} and~\ref{sec:pf} 
in hand, we are ready to calculate, and in Sections~\ref{sec:weak}, 
\ref{sec:strong} and~\ref{sec:conifold} we study the partition function and Wilson loop 
observables in the three natural limits of the matrix model. First, in Sections~\ref{sec:weak} 
we look at the orbifold point, which is the weak coupling point of the matrix model and likewise 
of the physical ABJM theory. The calculations there are straight-forward and we present 
the answers to these quantities. A single term ($1/6$ BPS loop at 2-loops) was calculated 
independently directly in the field theory. All other terms are predictions for the 
higher order perturbative corrections.

Section~\ref{sec:strong} addresses the strong coupling limit of the theory, where the 
matrix model should reproduce the semiclassical expansion of these observables in 
type IIA string theory on AdS$_4\times\IC\IP^3$. The expectation value of the 
Wilson loop was already derived in \cite{mp} and matched with a classical string in 
AdS. We first generalize the strong coupling expansion for the case of 
$N_1\neq N_2$, which corresponds to turning on a B-field in the AdS dual. 
This version of the theory 
was studied in \cite{abj} and a more precise analysis of the dictionary, capturing 
shifts in the charges, was presented in \cite{bergman,ah}. Interestingly, it turns out 
that the matrix model knows about these shifted charges, and the strong coupling 
parameter turns out to be exactly the one calculated in \cite{ah}, rather than the 
naive coupling.

In the same section we present also the calculation of the free energy in the matrix 
model. The result is proportional to $N^2/\sqrt\lambda$ (or a slight generalization 
for $N_1\neq N_2$). This scales at large $N$ like $N^{3/2}$, which is indeed 
the M-theory prediction for the number of degrees of freedom on $N$ coincident 
M2-branes \cite{klebanov}. Comparing with a supergravity calculation, we find precise agreement 
with the classical action of AdS$_4\times\IC\IP^3$. This is the first derivation of this 
large $N$ scaling in the field 
theory side. The matrix model also provides 
an infinite series of instanton/anti--instanton corrections to both the partition 
function and to the Wilson loop expectation value, which we interpret as fundamental strings 
wrapping the $\IC\IP^1$ inside $\IC\IP^3$. 

We then turn to a third limit of the theory, when one of the gauge couplings 
is perturbative and the other one not. In the strict limit the ABJM theory reduces 
to topological CS and in the matrix model one cut is removed. We show how 
to perform explicit calculations in this regime both from the planar solution of the 
matrix model and directly by performing matrix integrals. In both approaches 
one can see the full lens space matrix model arising as a (rather complicated) 
observable in topological CS theory on $\IS^3$. We speculate on possible tools 
of calculating directly in ABJM theory in this limit, where integrating out the 
bi-fundamental matter fields leads to correlation functions of Wilson loops in 
CS theory. We demonstrate the idea in the case of the $1/6$ BPS Wilson loop, 
which has a relatively simple perturbative expansion. This limit of the spin--chain 
of ABJM theory was considered in \cite{mss}, and a similar system in four 
dimensions was studied in \cite{rastelli}.

The brave souls that will make it to Sections~\ref{sec:modular} and~\ref{sec:wilson} 
will find some new results on the non-planar corrections to the matrix model, 
and hence to ABJM theory. In section 7 we show that 
the full $1/N$ expansion of the free energy on $\IS^3$ is completely determined by a recursive procedure based on 
direct integration \cite{hk,gkmw} of the holomorphic anomaly equations \cite{bcov}. The ability to determine 
the full expansion is closely related to the 
integrability of topological string theory on toric Calabi--Yau threefolds (as discussed in for example \cite{hkr}). By the AdS/CFT correspondence, 
the $1/N$ expansion obtained in this way determines the partition function of type IIA theory on the AdS$_4 \times \IC \IP^3$ background at all genera. This result 
is reminiscent of the ``old" matrix models for non-critical strings, where a double-scaled $1/N$ expansion, encoded in an integrable 
system, captures the all-genus partition function of a string theory. 
The recursive procedure for the computation of the $1/N$ expansion 
is quite efficient in practice, and one can perform explicit computations at high genus. This allows us to study the 
large genus behavior of the $1/N$ corrections, and we check that they display the factorial growth 
$\sim(2g)!$ typical of string perturbation theory \cite{shenker}. 
A careful examination of the coefficients suggests that this $1/N$ expansion is 
Borel summable. 

In Section~\ref{sec:wilson} we present the genus one correction to the Wilson 
loop and expand it at both weak and strong coupling. Another topic covered there 
is that of ``giant Wilson loops'' \cite{df,gpone,gptwo}, where in the supergravity 
dual (at least in AdS$_5\times\IS^5$) a fundamental string is replaced by a 
D-brane. This happens for Wilson loops in representations of dimension 
comparable to~$N$. We calculate the corresponding object in the matrix model 
and compare it to the vortex loop operators of \cite{vortex}.

One point we have not touched upon is the connection to topological strings. 
Since CS and the matrix model are related to topological strings, we expect there 
to be a direct connection between ABJM theory and a topological string theory. 
All the quantities captured by the matrix model should exist also in a topologically 
twisted version of ABJM theory, possibly along the lines of \cite{ks}.

\sectiono{The ABJM matrix model and Wilson loops}
\label{sec:MM}

\subsection{The matrix model and its planar limit}

The ABJM matrix model, obtained in \cite{kapustin}, gives an explicit integral expression for the partition function of the ABJM theory on $\IS^3$, as well as for 
Wilson loop VEVs. This matrix model is defined by the partition function 
\be
\label{kapmm}
\ba
&Z_\text{ABJM}(N_1, N_2, g_s)\\
&={\ri^{-\frac{1}{2}(N_1^2-N_2^2)}\over N_1! N_2!} \int \prod_{i=1}^{N_1}{ \rd \mu_i  \over 2\pi} \prod_{j=1}^{N_2} {\rd \nu_j \over 2\pi}
 {\prod_{i<j} \left( 2 \sinh \left( {\mu_i -\mu_j \over 2}\right) \right)^2 \left(2 \sinh \left( {\nu_i -\nu_j \over 2}\right) \right)^2 \over 
\prod_{i,j}  \left(2 \cosh \left( {\mu_i -\nu_j \over 2}\right) \right)^2} \re^{-{1\over 2g_s}\left(  \sum_i \mu_i^2 -\sum_j \nu_j^2\right)}, 
\ea
\ee
where the coupling $g_s$ is related to the Chern--Simons coupling $k$ of the ABJM theory as
\be
g_s={2 \pi \ri \over k}.
\label{gs}
\ee
In writing this matrix integral we have been very careful with its precise overall normalization, 
since one of our goals in the present paper is to compute the free energy on the sphere 
at strong coupling. The calculation of \cite{kapustin} captures the full $k$ dependence of the partition function, but we have to fix an overall 
$k$-independent normalization. This is done in two steps. First, we require that the above matrix integral 
reduces to the partition function for Chern--Simons theory 
on $\IS^3$ when $N_1=0$ or $N_2=0$ (in a specific framing of $\IS^3$). Once this is done, there is still a $k$-independent normalization factor which appears as a constant 
coefficient multiplying the $\cosh$ in the denominator. This term was not fixed in \cite{kapustin}, but it can be easily obtained from the formulae they presented. 
This calculation can be found in Appendix~\ref{sec:normalize}, 
and leads to the matrix integral (\ref{kapmm}).

The ABJM matrix model is closely related to the $L(2,1)$ lens space matrix model introduced in \cite{mm,akmv}. This matrix model is defined by the 
partition function
\be
\label{intdef}
\ba
Z_{L(2,1)}(N_1, N_2, g_s)={\ri^{-\frac{1}{2}(N_1^2+N_2^2)}\over N_1! N_2!} \int \prod_{i=1}^{N_1}{ \rd \mu_i  \over 2\pi} \prod_{j=1}^{N_2} {\rd \nu_j \over 2\pi} & \prod_{i<j} \left( 2 \sinh  \left( {\mu_i -\mu_j \over 2}\right) \right)^2 \left( 2 \sinh  \left( {\nu_i -\nu_j \over 2}\right) \right)^2\\
\times &\prod_{i,j} \left( 2  \cosh  \left( {\mu_i -\nu_j \over 2}\right) \right)^2\,  \re^{-{1\over 2g_s}\left(  \sum_i \mu_i^2 +\sum_j \nu_j^2\right)}.
\ea
\ee
The relation between the partition functions is simply \cite{mp}
\be
\label{changesignZ}
Z_\text{ABJM}(N_1, N_2, g_s) =Z_{L(2,1)}(N_1, -N_2, g_s). 
\ee
Since the large $N$ expansion of the free energy gives a sequence of analytic functions of $N_1$, $N_2$, once these functions are known in one model, they can be obtained in the other 
by the trivial change of sign $N_2 \rightarrow -N_2$.

\FIGURE{
\includegraphics[height=4.5cm]{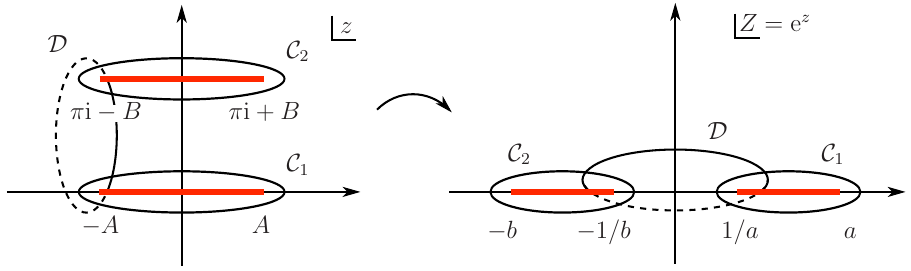} 
\caption{Cuts in the $z$-plane and in the $Z$-plane.}
\label{cuts}
}

Let us now discuss the large $N$ solution of the lens space matrix model, following 
\cite{akmv,hy,mp}. At large $N$, the two sets of eigenvalues, $\mu_i$, $\nu_j$, condense 
around two cuts. The cut of the $\mu_i$ eigenvalues is centered around $z=0$, while that of the 
$\nu_i$ eigenvalues is centered around $z=\pi \ri$. We will write the cuts as
\be
\label{ABcuts}
\CC_1=(-A, A), \qquad \CC_2=(\pi \ri -B, \pi \ri +B),
\ee
in terms of the endpoints $A, B$. It is also useful to use the exponentiated variable 
\be
Z=\re^z, 
\ee
In the $Z$ plane the cuts (\ref{ABcuts}) get mapped to 
\be
\label{Zend}
(1/a, a), \qquad (-1/b, -b) , 
\qquad\qquad
a=\re^A, \qquad b =\re^B, 
\ee
which are centered around $Z=1$, $Z=-1$, respectively, see \figref{cuts}. 
We will use the same notation $\CC_{1,2}$ for the cuts in the $Z$ plane. The 
large $N$ solution is encoded in the total resolvent of the matrix model, 
$\omega(z)$. It is defined as \cite{hy}
\be
\label{resolv}
\omega(z)
=g_s\left<\tr\left(\frac{Z+U}{Z-U}\right)\right>
=g_s \left\langle \sum_{i=1}^{N_1} \coth \left( {z-\mu_i \over 2} \right) \right\rangle +g_s \left\langle\sum_{j=1}^{N_2}  \tanh \left( {z-\nu_j \over 2} \right)\right\rangle
\ee
where
\be
\label{U}
U=\begin{pmatrix}\re^{\mu_i} &0 \\ 0& -\re^{\nu_j} \end{pmatrix}.
\ee
We will denote by $\omega_0(z)$ the planar limit of the resolvent, which was found in explicit form in \cite{hy}. It reads, 
\be
\label{explicitRes}
\omega_0(z) =2 \log \biggl( {\re^{-t/2} \over 2} \Bigl[ {\sqrt{(Z+b)(Z+1/b)}}-{\sqrt{(Z-a)(Z-1/a)}}\Bigr]\biggr),  
\ee
where 
\be
t=t_1+t_2
\ee
is the total 't~Hooft parameter. It is useful to introduce the variables\footnote{The variable $\beta$ is related to the variable $\xi$ in \cite{mp} by $\beta=\xi/2$.} 
\be
\label{zetadef}
\zeta={1\over 2}\left( a  +{1\over a} -b -{1\over b}\right), 
\qquad 
\beta={1 \over 4}\left( a  +{1\over a} +b +{1\over b}\right).
\ee
$\beta$ is related to the total 't~Hooft parameter through
\be
\label{gthooft}
\beta=  \re^{t}. 
\ee
All the relevant planar quantities can be expressed in terms of period integrals of the one-form $\omega_0(z)\rd z$. The 't~Hooft parameters 
are given by
\be
\label{tperiods}
t_i ={1\over 4\pi \ri} \oint_{\CC_i} \omega_0 (z)  \rd z , \qquad i=1,2.
\ee
The planar free energy $F_0$ satisfies the equation
\be
\label{pder}
\CI\equiv
{\partial F_0 \over \partial t_1} -{\partial F_0 \over \partial t_2}-\frac{\pi\ri t}{2}
=-\frac{1}{2}\oint_{{\cal D}} \omega_0(z) \rd z, 
\ee
where the ${\cal D}$ cycle encloses, in the $Z$ plane, the interval between $-1/b$ 
and $1/a$ (see \figref{cuts}).%
\footnote{Likewise one can calculate the second ``B-cycle'' period, and it will arise when 
solving the Picard-Fuchs equations at strong coupling in Section~\ref{sec:pf-strong}.}

The derivatives of these periods can be calculated in closed form by adapting a trick from \cite{bt}. One finds, 
\be
\label{derperzeta}
\frac{\d t_{1,2}}{\d\zeta}=-
\frac{1}{4\pi \ri}\oint\limits_{\mathcal{C}_{1,2}}\frac{\rd Z}{\sqrt{(Z^2-
\zeta Z+1)^2-4\beta^2 Z^2}}=\pm \frac{\sqrt{ab}}{\pi(1+ab)}\,K(k),
\ee
and similarly
\be
\label{derperbeta}
 \frac{\d t_1}{\d\beta}
 =-2 \frac{\sqrt{ab}}{\pi(1+ab)}\left(K(k)-\frac{2ab}{1+ab}\,\Pi(n_1|k)-\frac{2}{1+ab}\,\Pi(n_2|k)\right),
\end{equation} 
where
\begin{equation}
 k^2=1-\left(\frac{a + b}{1 + ab}\right)^2, \qquad 
 n_1=\frac{1-a^2}{1+ab}, \qquad 
 n_2=\frac{b(a^2-1)}{a(1+ab)}\,.
\end{equation} 
Likewise for the period integral in (\ref{pder}) we find 
\begin{equation}
\ba
 \frac{\d \CI}{\d\zeta}&=
 -2\,\frac{\sqrt{ab}}{1+ab}\,K(k'), \\
  \frac{\d \CI}{\d\beta}&=
 4\,\frac{\sqrt{ab}}{1+ab}\left(K(k')+\frac{2a(1 - b^2)}{(1 + ab)(a + b)}(\Pi(n_1'|k')-\Pi(n_2'|k'))\right),
 \ea
\label{I-derivs}
\end{equation} 
where
\begin{equation}
 k'=\frac{a + b}{1 + ab}, \qquad  
 n_1'=\frac{a + b}{b(1 + ab)}, \qquad 
 n_2'=\frac{b(a + b)}{1 + ab}.
\end{equation} 

We can now use the dictionary between the lens space matrix model and the ABJM 
matrix model given by (\ref{gs}) and (\ref{changesignZ}) to get the planar solution of 
the latter model. In particular, the natural 't~Hooft parameters in the ABJM theory
\be
\lambda_j={N_j\over k}
\ee
are obtained from the planar solution of the lens space matrix model by the replacement 
\be
\label{relthoofts}
t_1=2\pi \ri \lambda_1, \qquad t_2=-2\pi \ri \lambda_2. 
\ee
Since in the ABJM theory the couplings $\lambda_{1,2}$ are real, the matrix model 
couplings $t_{1,2}$ are pure imaginary. Thanks to (\ref{gthooft}) we know that 
$\beta$ is of the form
\be
\label{betalambda}
\beta= \re^{2\pi \ri (\lambda_1 -\lambda_2)} 
\ee
{\em i.e.}, it must be a phase.

For later convenience we introduce yet another parameterization of the couplings in 
terms of $B$ and $\kappa$
\be
B =\lambda_1 -\lambda_2 +{1\over 2},\qquad
\kappa=\re^{-\pi \ri B}\zeta\,.
\label{zetakappa}
\ee
$B$ is identified as the B-field in the dual type IIA background \cite{ah}. Notice that it 
has a shift by $-1/2$ as compared to the original prescription in \cite{abj}. Clearly, 
all calculations in the matrix model are periodic under $B\to B+1$, up to possible 
monodromies (see (\ref{monodromy}) below). As we shall 
see later, the parameter $\kappa$ is real for physical values of $\lambda_{1,2}$.

\subsection{Wilson loops}
\label{sec:wl-intro}

One of the main results of \cite{kapustin} is that the VEV of the $1/6$ BPS Wilson loop in ABJM theory, labelled by a representation $R$ or $U(N_1)$, 
can be obtained by calculating the VEV of the matrix $\re^{\mu_i}$ 
in the matrix model (\ref{kapmm}), {\em i.e.}, 
\be
\label{16WL}
\langle W^{1/6}_R\rangle
=g_s \left\langle \tr_R\left (\re^{\mu_i}\right)  \right\rangle_\text{ABJM MM},
\ee

A 1/2 BPS loop $W^{1/2}_\CR$ was constructed in \cite{dt} , where $\CR$ is a 
representation of the supergroup $U(N_1|N_2)$. in \cite{dt} it was also shown 
that it localizes to the matrix model correlator in the ABJM matrix model
\be
\label{12wl}
\langle W^{1/2}_\CR \rangle
=g_s \left\langle \text{Str}_\CR\, U \right\rangle_\text{ABJM MM},
\ee
with the same $U$ as in (\ref{U}). Though at first sight the minus sign on the lower 
block of $U$, may look surprising, it can be 
attributed to the fact that the $\nu_j$ eigenvalues are shifted by $\pi\ri$ from the 
real line. Due to the relation between the ABJM matrix 
model and the lens space matrix model, these correlators can be computed in the 
lens space matrix model as follows:
\be
\ba
\langle W^{1/6}_R\rangle &= g_s\left\langle \tr_R\left (\re^{\mu_i}\right)  \right\rangle_{L(2,1)} \Big|_{N_2 \rightarrow -N_2}, \\
\langle W^{1/2}_\CR \rangle&=g_s \left\langle \tr_\CR \,U \right\rangle_{L(2,1)}\Big|_{N_2 \rightarrow -N_2},
\ea
\label{L21-wl}
\ee
where the super-representation $\CR$ is regarded as a representation of $U(N_1+N_2)$. 

To evaluate the Wilson loop one uses the resolvent, or equivalently, the eigenvalue densities
\be
\label{gdensities}
\ba
\rho^{(1)}(Z) \rd Z 
&=-{1\over 4 \pi \ri t_1}{\rd Z \over Z} \left( \omega(Z+\ri \epsilon) -\omega(Z-\ri \epsilon)\right), 
\qquad &Z&\in \CC_1,\\
\rho^{(2)}(Z) \rd Z 
&={1\over 4 \pi \ri t_2 } {\rd Z \over Z}\left( \omega(Z+\ri \epsilon) -\omega(Z-\ri \epsilon)\right), 
\qquad &Z&\in \CC_2.
\ea
\ee
which are each normalized in the planar approximation to unity
\be
\int_{\CC_i}\rho_0^{(i)}\,\rd Z=1.
\ee
For the $1/6$ BPS Wilson loop in the fundamental representation one needs to 
integrate $\re^z=Z$ over the first cut
\be
\left \langle W^{1/6}_{\tableau{1}} \right\rangle
=t_1 \int_{\CC_1} \rho^{(1)} (Z) Z \rd Z
=\oint_{\CC_1}\frac{\rd Z}{4\pi \ri}\,\omega(Z).
\label{1/6-integral}
\ee
The correlator relevant for the $1/2$ BPS Wilson loop (again in the fundamental representation) is much easier, since
\be
\left \langle W^{1/2}_{\tableau{1}} \right\rangle
=t_1 \int_{\CC_1} \rho^{(1)} (Z) Z \rd Z - 
t_2 \int_{\CC_2} \rho^{(2)} (Z) Z \rd Z
= \oint_{\infty} {\rd  Z \over 4 \pi \ri }\, \omega(Z) 
\label{1/2-integral}
\ee
and it can be obtained by expanding $\omega(Z)$ around $Z \to \infty$. 

The comparison to the case of the $1/2$ BPS Wilson loop in $\CN=4$ SYM in 4d 
is straight-forward. In that case the matrix model is Gaussian and 
the eigenvalue density in the planar approximation follows 
Wigner's semi-circle law. Doing the 
integral with the insertion of $\re^z$ gives a modified Bessel function 
\cite{esz}
\be
\rho_0(z)=\frac{2}{\pi\lambda}\sqrt{\lambda-z^2}
\quad\Rightarrow\quad
\langle W^{1/2}_\text{4d $\CN=4$}\rangle_\text{planar}
=\int_{-\sqrt\lambda}^{\sqrt\lambda}\rho_0(z)\,\re^z\,\rd z
=\frac{2}{\sqrt{\lambda}}I_1(\sqrt\lambda).
\label{N=4wl}
\ee

For the ABJM matrix model all the expressions are more complicated. Still the 
derivative with respect to $\zeta$ and $\beta$ of the integral expression for the 
$1/6$ BPS Wilson loop (\ref{1/6-integral}) can be written in closed form \cite{mp}, 
like the integrals (\ref{derperzeta}) and (\ref{derperbeta})
\be
\ba
\partial_\zeta\langle W^{1/6}_{\tableau{1}} \rangle
&=-{1\over \pi} \frac{1}{\sqrt{ab}(1+ab)}\left(a\, K(k)-(a+b)\,\Pi(n|k)\right)
\\
\partial_\beta\langle W^{1/6}_{\tableau{1}} \rangle
&=-{2 \over \pi} \frac{\sqrt{ab}}{a+b}\,E(k)\,.
\ea
\label{16-explicit}
\ee
For the $1/2$ BPS Wilson loop of \cite{dt} the situation is much simpler and 
in the planar approximation one needs only the 
large $Z$ behavior of $\omega_0$ (\ref{explicitRes})
\be
\omega_0=t+\frac{\zeta }{Z}+\frac{\zeta ^2+2\beta ^2-2}{2Z^2}+
\frac{\zeta(\zeta^2+6\beta^2-3)}{3 Z^3}+\CO(Z^{-4}).
\label{omega-expand}
\ee
One finds \cite{mp} 
\be
\label{12planar}
\langle W^{1/2}_{\tableau{1}} \rangle_\text{planar}={\zeta \over 2},
\ee
which can then be expanded in different regimes. We will elaborate on the expansion 
of this expression in the next sections and will also turn to the non-planar corrections 
to it and to that of the $1/6$ BPS loop in Section~\ref{sec:wilson}.

As a simple generalization, by the replacement $Z\to Z^l$ on the right hand side of 
(\ref{1/2-integral}), the 
higher order terms in the expansion (\ref{omega-expand}) give the expectation 
values of multiply wrapped $1/2$ BPS Wilson loops where $U\to U^l$ in (\ref{12wl}). 
For even winding the sign in the lower block of the matrix $U$ (\ref{U}) is absent. This 
is consistent with the gauge theory calculation \cite{dt}, where this sign arose 
from the requirement of supersymmetry invariance in the presence of the fermionic 
couplings which are antiperiodic, as should be the case for a singly-wound 
contractible cycle (see also the discussion in \cite{ll}).

The normalization of the Wilson loop as given by (\ref{1/6-integral}) and 
(\ref{1/2-integral}) is not the same as in the 4d $\CN=4$ case (\ref{N=4wl}). For the 
$1/6$ BPS loop, the leading term at weak coupling is $t_1=2\pi\ri N_1/k$. This means 
that the trace in the fundamental is normalized by a factor of $g_s$. For the $1/2$ 
BPS loop the leading term is $t_1\pm t_2=g_s(N_1\mp N_2)$, where the sign 
depends on the winding number. We will comment more about this normalization 
in Section~\ref{sec:WL-strong}.

\sectiono{Moduli space, Picard--Fuchs equations and periods}
\label{sec:pf}

In this section we present the tools for solving the lens space matrix model using 
special geometry. We present three special points in the moduli space of the theory 
and write explicit expressions for the four periods of $\omega_0$ at the vicinity 
of these points.

The lens space matrix model is equivalent to topological string theory on 
local $\IF_0=\IP^1 \times \IP^1$. The $1/N$ expansions of the free energy and of the
1/2 BPS Wilson loop VEV are the genus expansions of closed and open topological 
string amplitudes. The planar content of the theory is encoded in the periods of the 
mirror geometry described by the family of elliptic curves $\Sigma$, which can be written as 
\be
\label{mc}
y={z_1 x^2 + x +1 -{\sqrt{(1+x + z_1 x^2)^2-4 z_2 x^2}}\over 2}. 
\ee
Here, $z_1, z_2$ parametrize the moduli space of complex structures, which is the mirror to the enlarged K\"ahler moduli space of 
local $\IF_0$. This moduli space has a very rich structure first uncovered in \cite{akmv} and further studied in, for example, \cite{bt,hkr} by using the standard techniques of 
mirror symmetry. 

Notice that the mirror geometry (\ref{mc}) is closely related to 
the resolvent $\omega_0(Z)$. Indeed, one finds that $\omega_0(Z)\sim \log \, y(x)$ provided we identify the variables as
\be
x=-Z z_1^{-1/2}, 
\ee
and 
\be
\label{zxi}
\zeta={1\over {\sqrt{ z_1}}}, \qquad \beta= {\sqrt{z_2\over z_1}}.
\ee
This can also be expressed as (\ref{zetakappa})
\be
z_1=\frac{\re^{-2\pi\ri B}}{\kappa^2},
\qquad
z_2=\frac{\re^{2\pi\ri B}}{\kappa^2}.
\label{z1z2-Bkappa}
\ee

Let us now discuss in some detail the moduli space of (\ref{mc}), since it will play a fundamental role in the following. It has complex dimension two, corresponding to the two  
complexified K\"ahler parameters of local $\IF_0$. The coordinates $z_1, z_2$ (or $\zeta, \beta$) are global coordinates in this moduli space. Another way of parametrizing it is to use 
the periods of the meromorphic one-form 
\be
\omega=\log \, y(x) {\rd x \over x}
\ee
As it is well-known, these periods are annihilated by a pair of differential operators called Picard--Fuchs operators. In terms of $z_1, z_2$, the operators are 
\be
\label{pfsystem}
\ba
{\cal L}_1 &= z_2 (1-4 z_2) \xi^2_2 - 4 z_1^2 \xi_1^2 -
8 z_1 z_2 \xi_1 \xi_2 - 6 z_1\xi_1 +(1- 6 z_2) \xi_2, \\
{\cal L}_2 &= z_1 (1-4 z_1) \xi^2_1 - 4 z_2^2 \xi_2^2 -
8 z_1 z_2 \xi_1 \xi_2 - 6 z_2\xi_2 +  (1-6 z_1)\xi_1, 
\ea
\ee
where 
\be
\xi_i={\partial\over \partial z_i}.
\ee
These operators lead to a system of differential equations known as {\it Picard--Fuchs} (PF) equations. An important property of the moduli space is the existence of 
special points, generalizing the regular singular points of ODEs on $\IC$. The PF system can be solved around these points, and the solutions give a basis for the periods of the meromorphic one-form. We can use two of the solutions to parametrize the moduli space near a singular point, and the resulting local coordinates, given by periods, are 
usually called {\it flat coordinates}. 

\subsection{Orbifold point, or weak coupling}
There are three types of special points in the moduli space. The first one is the {\it orbifold} point discovered in \cite{akmv}, which is 
the relevant one in order to make contact with the matrix model. 
To study this point one has to use the global variables
\be \label{cov} 
 x_1=1-{z_1\over z_2}, \qquad x_2={1\over \sqrt{z_2}\left(1-{z_1\over z_2}\right)}.
\ee
The orbifold point is then defined as $x_1=x_2=0$, and in terms of these variables the Picard--Fuchs system is given by the two operators
\be
\ba
 \mathcal{L}_1=&\,\frac{1}{4} (8-8 x_1+x_1^2) x_2 \d_{x_2}
 -\frac{1}{4}\left(4-(2-x_1)^2 x_2^2\right) \d^2_{x_2}
-x_1 (2-3 x_1+x_1^2) x_2 \d_{x_1}\d_{x_2} \\& 
 -(1-x_1) x_1^2 \d_{x_1}+(1-x_1)^2 x_1^2 \d^2_{x_1},\\
 \mathcal{L}_2=&\,(2-x_1) x_2 \d_{x_2}-(1-(1-x_1) x_2^2) \d^2_{x_2}-x_1^2 \d_{x_1}
 -2 (1-x_1) x_1 x_2 \d_{x_1}\d_{x_2}+(1-x_1) x_1^2 \d_{x_1}^2.
 \ea
\ee
A basis of periods near the orbifold point was found in \cite{akmv}. It reads, 
\be
\label{speriods}
\ba
\sigma_1&=-\log(1-x_1),\\
\sigma_2&=\sum_{m,n} c_{m,n} x_1^m x_2^n,\\
\CF_{\sigma_2}&=\sigma_2 \log x_1+\sum_{m,n} d_{m,n} x_1^m x_2^n \ ,
\ea
\ee
where the coefficients $c_{m,n}$ and $d_{m,n}$ vanish for non-positive $n$ or $m$ 
as well as for all even $n$. They satisfy the following recursion relations with the 
seed values $c_{1,1}=1$, $d_{1,1}=0$ and $d_{1,3}=-1/6$: 
\be
\label{cdrecursion}
\ba
c_{m,n}&={(n+2-2m)^2\over 4 (m-n) (m-1)}\,c_{m-1,n},\cr
c_{m,n}&=\frac{(n-2)^2 (m-n+2) (m-n+1)}{n(n-1) (2m-n)^2}\,c_{m,n-2},\\
d_{m,n}&=\frac{(n+2-2 m)^3d_{m-1,n}+4(n^2-n-2 m+2)c_{m,n}}{4 (m-1) (m-n) (n+2-2 m)},\cr
d_{m,n}&=\frac{(n-2)^2 (m-n+1) (m-n+2)}{n(n-1) (2m-n)^2}\,d_{m,n-2} 
+ \left(\frac{1}{m-n+2}+\frac{1}{m-n+1}+\frac{4}{n-2 m}\right)c_{m,n}.
\ea
\ee
The 't~Hooft parameters of the matrix model are period integrals of the meromorphic one-form, therefore they must be linear combinations of the periods above, and one finds \cite{akmv}
\be
t_1={1\over 4} (\sigma_1+\sigma_2), \qquad
t_2={1\over 4} (\sigma_1-\sigma_2).
\label{sigma-t}
\ee
An expansion around the orbifold point leads to a regime in which $t_1, t_2$ are very small. In view of (\ref{relthoofts}) this corresponds, in the ABJM model, to the {\it weakly coupled theory}
\be
\lambda_1, \, \lambda_2 \ll 1. 
\ee
The remaining period in (\ref{speriods}) might be used to compute the genus zero free energy of the matrix model. Using 
the normalization appropriate for the ABJM matrix model, we find
\be
\label{I-weak}
\CI=4{\partial F_0 \over \partial \sigma_2}-\frac{\pi\ri t}{2}
= {1\over 2} \CF_{\sigma_2} -\log(4) \sigma_2-\frac{\pi\ri}{2}\sigma_1. 
\ee

\subsection{Large radius, or strong coupling} 
\label{sec:pf-strong}

The second point that we will be interested in is the so-called {\it large radius point} corresponding to $z_1=z_2=0$. 
This is the point where the Calabi--Yau manifold is in its 
geometric phase, and the expansion of the genus zero free energy near that point leads to the counting of holomorphic curves with Gromov--Witten invariants. The solutions 
to the Picard--Fuchs equations (\ref{pfsystem}) near this point can be obtained in a systematic way by considering the so-called fundamental period 
\be
\varpi_0(z_1, z_2; \rho_1, \rho_2) =\sum_{k,l\geq0} 
\frac{\Gamma(2k+2l + 2\rho_1 + 2\rho_2) \Gamma(1+\rho_1)^2\,\Gamma(1+\rho_2)^2}
{\Gamma(2\rho_1+2\rho_2)\Gamma(1+ k +\rho_1)^2\,\Gamma(1+ l +\rho_1)^2}z_1^{k+\rho_1}z_2^{l+\rho_2}.
\ee
As reviewed in for example \cite{hkt}, a basis of solutions to the PF equations can be obtained by acting on the fundamental period with the following 
differential operators
\be
D_i^{(1)}= \partial_{\rho_i}, \qquad D_i^{(2)}={1\over 2} \kappa_{ijk} \partial_{\rho_j}\partial_{\rho_k}.
\ee
Here $\kappa_{ijk}$ are the classical triple intersection numbers of the Calabi--Yau. This leads to the periods
\be
\ba
T_i(z_1, z_2)&=-D_i^{(1)}\varpi_0(z_1, z_2; \rho_1, \rho_2)\Bigl|_{\rho_1=\rho_2=0}, \\
F_i(z_1, z_2)&=-D_i^{(2)}\varpi_0(z_1, z_2; \rho_1, \rho_2)\Bigl|_{\rho_1=\rho_2=0}.
\ea
\ee
These periods should be linearly related to those defined in the matrix model in equations 
(\ref{tperiods}) and (\ref{pder}). We present now some explicit expressions for them that 
we will use in Sections~\ref{sec:analytic} and~\ref{sec:free} to solve for these relations 
(see equations (\ref{tFFT}) and (\ref{freefirst})).

In general, one normalizes these periods and divides them by the fundamental period evaluated at $\rho_1=\rho_2=0$. But in local mirror symmetry we have \cite{local}
\be
\varpi_0(z_1, z_2; \rho_1, \rho_2)\Bigl|_{\rho_1=\rho_2=0}=1. 
\ee
The $T_i$ are single-logarithm solutions, and they are identified in standard mirror symmetry with the complexified K\"ahler parameters, while the $F_i$ are double-logarithm solutions 
and they are identified with the derivatives of the large radius genus zero free energy w.r.t. the $T_i$. In our case, we find the explicit expressions   
\be
\label{mirrormapone}
\ba
-T_1&= \log z_1 + \omega^{(1)} (z_1, z_2),\\
-T_2&= \log z_2 + \omega^{(1)} (z_1, z_2),
\ea
\ee
where 
\be
\omega^{(1)} (z_1, z_2) = 2\sum_{k,l\ge 0, \atop (k,l)\not=(0,0)} { \Gamma(2k + 2l) \over \Gamma(1+k)^2 \Gamma(1+l)^2} z_1^k z_2^l =2z_1 + 2z_2 + 3 z_1^2 + 12 z_1 z_2 + 3 z_2^2 + \cdots
\ee
In order to obtain the $F_i$ we have to compute the double derivatives w.r.t. the parameters $\rho_1$, $\rho_2$. We find
\be
\partial_{\rho_1}^2\varpi_0(z_1, z_2; \rho_1, \rho_2)\Bigl|_{\rho_1=\rho_2=0} = \log^2 z_1 + 2 \log z_1 \, \omega^{(1)} (z_1, z_2) +  \omega^{(2)}_1(z_1, z_2),
\ee
where
\be
\omega^{(2)}_1(z_1, z_2)=8 \sum_{k,l\ge 0, \atop (k,l)\not=(0,0)} { \Gamma(2k + 2l) \over \Gamma(1+k)^2 \Gamma(1+l)^2} \left( \psi(2k+2l) -\psi(1+k)\right)  z_1^k z_2^l.
\ee
Similarly, 
\be
\partial_{\rho_2}^2\varpi_0(z_1, z_2; \rho_1, \rho_2)\Bigl|_{\rho_1=\rho_2=0} = \log^2 z_2 + 2 \log z_2 \, \omega^{(1)} (z_1, z_2) + \omega^{(2)}_2 (z_1, z_2) 
\ee
where
\be
\omega^{(2)}_2(z_1, z_2)
=8 \sum_{k,l\ge 0, \atop (k,l)\not=(0,0)} { \Gamma(2k + 2l) \over \Gamma(1+k)^2 \Gamma(1+l)^2} \left( \psi(2k+2l) -\psi(1+l)\right) z_1^k z_2^l 
=\omega^{(2)}_1(z_2, z_1).
\ee
Finally, 
\be
\ba
\partial_{\rho_1}\partial_{\rho_2}\varpi_0(z_1, z_2; \rho_1, \rho_2)\Bigl|_{\rho_1=\rho_2=0} 
=&\,\log z_1 \log z_2 + \left( \log z_1 + \log z_2\right) \omega^{(1)} (z_1, z_2) \\
&+ {1\over 2} \left( \omega^{(2)}_1(z_1, z_2)+ \omega^{(2)}_2(z_1, z_2) \right). 
\ea
\ee
The double log periods are obtained as linear combinations of the above, by using the explicit expressions for the classical intersection numbers that can be found in for 
example \cite{hkr}
\be
\kappa_{111}={1\over 4}, \qquad \kappa_{112}=-{1\over 4}, \qquad \kappa_{122}=-{1\over 4}, \qquad\kappa_{222}={1\over 4}.
\ee
We find: 
\be
\label{dperiods}
\ba
F_1(z_1, z_2)&=- {1\over 8} \left( \partial_{\rho_1}^2 \varpi_0 -2\partial_{\rho_1}\partial_{ \rho_2} \varpi_0 -\partial_{\rho_2}^2 \varpi_0 \right) \\
&=-{1\over 8}\left( \log^2 z_1  -2 \log z_1\log z_2 -\log^2 z_2 \right) +{1\over 2} \log z_2\, \omega^{(1)} (z_1, z_2) + {1\over 4} \omega^{(2)}_2 (z_1, z_2), \\
F_2 (z_1, z_2)&=- {1\over 8} \left( -\partial_{\rho_1}^2 \varpi_0 -2\partial_{\rho_1} \partial_{\rho_2} \varpi_0 +\partial_{\rho_2}^2 \varpi_0  \right) \\
&=-{1\over 8}\left(- \log^2 z_1  -2 \log z_1\log z_2 +\log^2 z_2 \right) +{1\over 2} \log z_1\, \omega^{(1)} (z_1, z_2) +{1\over 4} \omega^{(2)}_1 (z_1, z_2).
\ea
\ee
They satisfy the symmetry property
\be
F_1(z_1, z_2)=F_2(z_2, z_1). 
\ee

The reason why we are interested in the large radius point is because it describes the structure of the ABJM theory at {\it strong coupling}. In the region where $z_2$ is small, $x_2$ is large and the periods $t_{1,2}$ grow. In general, the expansions of the periods 
around the special points have a finite radius of convergence, but they can be analytically continued to the other ``patches". Since their analytic continuation satisfies the 
PF equation, we know for example that the analytic continuation of the orbifold periods to the large radius patch must be linear combinations of the periods at large radius. 
This provides an easy way to perform the analytic continuation which will be carried out in detail in the Section~\ref{sec:strong}, where we will verify that indeed the region near the large radius point corresponds to 
\be
\lambda_1, \, \lambda_2 \gg 1.
\ee

\subsection{Conifold locus}

Finally, the third set of special points is the {\it conifold locus}. This is defined by $\Delta=0$, where
\be
\Delta=1-8(z_1+z_2)+16 (z_1-z_2)^2.
\ee
In terms of the variables $\zeta, \beta$, this locus corresponds to the four lines
\be
\label{fourlines}
\zeta =-2 \beta \pm 2, \qquad \zeta=2\beta \pm 2. 
\ee
The conifold locus is the place where cycles in the geometry collapse to zero size. 
The first two lines correspond to $a=\pm 1$, {\em i.e.}, the collapse of the $\CC_1$ cycle, 
while the second set of lines corresponds to $b=\mp 1$, {\em i.e.}, to the collapse of the $\CC_2$ cycle. 
In principle we can solve the PF system near any point in the conifold locus, but in practice it is useful to focus on the point 
\be
\label{symconi}
z_1=z_2={1\over 16}
\ee
which has been studied in \cite{hkr}. We will call it the symmetric conifold point. Appropriate global coordinates around this point are\footnote{These are slightly different from the ones used in 
\cite{hkr}.}
\begin{equation}
\label{conglobal}
 y_1=1-\frac{z_1}{z_2}, \qquad y_2=1-\frac{1}{16z_1}.
\ee
In terms of these coordinates, the PF system reads
\begin{equation}
\label{pfconi}
\ba
 \mathcal{L}_1=&\,
 \d_{y_2}-2 (1-y_2) \d_{y_2}^2-8 (1-y_1)^2 \d_{y_1}+8 (1-y_1)^3 \d_{y_1}^2, \\
 \mathcal{L}_2=&\,
 -(7-8 y_2)\d_{y_2}+2 (3-7 y_2+4 y_2^2) \d_{y_2}^2-8 (1-y_1)  \d_{y_1}\\
& -16 (1-y_1) (1-y_2) \d_{y_1} \d_{y_2}+8 (1-y_1)^2  \d_{y_1}^2.
\ea
\ee
Notice that, strictly speaking, the orbifold point does not belong to the conifold locus, once the moduli space is compactified and resolved \cite{akmv}. A generic point in the 
conifold locus has then $t_1=0$ or $t_2=0$, but not both, and expanding around the conifold locus means, in the ABJM theory, an expansion in the region
\be
\lambda_1 \ll 1, \qquad \lambda_2 \sim1, 
\ee
or in the region with $\lambda_2$ exchanged with $\lambda_1$. This regime of the 
ABJM theory has been considered in \cite{mss}. 

It was observed in \cite{abk} that the moduli space of the local $\IF_0$ surface can be mapped to a well-known moduli space, 
namely the Seiberg--Witten (SW) $u$-plane \cite{sw}. This plane is parametrized by a single complex variable $u$. The relation between the moduli is 
\be
\label{umodulus}
u={1\over 2}\left( \beta + \beta^{-1} \right) -{\zeta^2 \over 8 \beta}. 
\ee
The three singular points that we have discussed (large radius, orbifold, and symmetric conifold) map to the points $u=\infty, +1, -1$. These are the semiclassical, monopole and dyon points 
of SW theory. As we will see, they can be identified with interesting points in ABJM theory.

An important set of quantities in the study of moduli spaces of CY threefolds are the three-point couplings or Yukawa couplings, $C_{z_i z_j z_k}$. 
These are the components of a completely symmetric 
degree three covariant tensor on the moduli space. When expressed in terms of flat coordinates they give the third derivatives of the genus zero free energy. 
In terms of the coordinates $z_1, z_2$, the Yukawa couplings are given by \cite{akmv,hkr}
\be
 \ba\label{yuka}
 C_{111}&={(1 - 4 z_2)^2 - 16 z_1 (1 + z_1) \over 4 z_1^3 \Delta}, \\
 C_{112}&={16 z_1^2 - (1 - 4 z_2)^2 \over 4 z_1^2 z_2 \Delta},\\
 C_{122}&= {16 z_2^2 - (1 - 4 z_1)^2 \over 4 z_1 z_2^2 \Delta},\\
 C_{222}&={(1 - 4 z_1)^2 - 16 z_2 (1 + z_2) \over 4 z_2^3 \Delta}.
 \ea
\ee

\subsection{The moduli space of the ABJM theory}

The matrix model of ABJM is closely related to the lens space matrix model, and 
therefore so are also the moduli spaces of the theories. Some of the explicit relations needed 
for this identification will be presented only in the following sections, but we would still like 
to present here the main points on the moduli space.

We can think about the {\it moduli space of the planar ABJM theory} as the space of admissible values of the 't~Hooft parameters $\lambda_1, \lambda_2$. We will assume for simplicity that $k>0$. The theory with negative values of $k$ can be obtained from this one by a parity transformation. In the gauge theory $\lambda_{1,2}$ must be rational and non-negative (for $k>0$). Moreover, according to \cite{abj}, any 
value of $\lambda_{1,2}$ is admissible as long as 
\be
|\lambda_1 -\lambda_2|\le 1. 
\ee
This moduli space can be parametrized by the $B$ field and $\kappa$, which from 
the explicit expressions derived below (\ref{weakkappa}) and (\ref{fullkappa}) 
has to be real and positive. It can be identified as a real submanifold of the moduli 
space of local $\IF_0$. Moreover, we can identify the singular points of this 
moduli space with natural limits of ABJM theory (see \figref{moduliABJMf}):

\FIGURE{
\includegraphics[height=7cm]{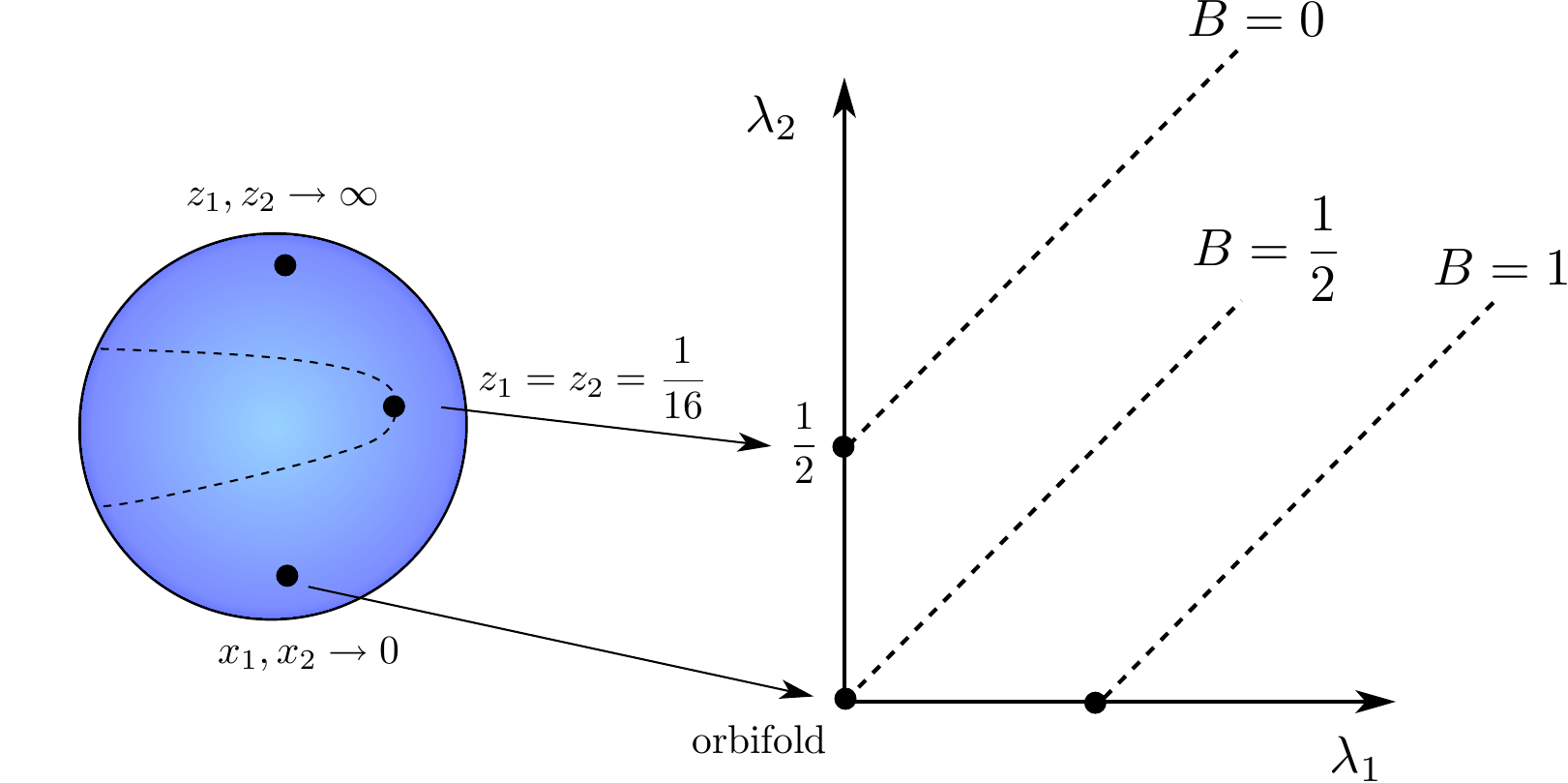} 
\caption{The moduli space of the ABJM theory, describing the possible values of the 't~Hooft couplings $\lambda_{1,2}$, can be parametrized by a real submanifold of 
the moduli space of local $\IF_0$, here depicted as a sphere. 
The orbifold point maps to the origin, while the conifold locus (which is represented by a dashed line) maps to the two axes.}
\label{moduliABJMf}
}

\begin{enumerate}

\item The {\it weak coupling regime} $\lambda_{1,2} \to 0$ corresponds to the orbifold 
point of the local $\IF_0$ geometry $\kappa=0$, $B=1/2$. In terms of type IIA theory, 
this is also an orbifold geometry with a small radius but a nonzero 
value for the $B$ field. 

\item The {\it strong coupling regime} $\lambda_{1,2} \to \infty$ (where also $\kappa\to\infty$) 
corresponds to the large radius limit of the local $\IF_0$ geometry. 

\item Out of the four lines (\ref{fourlines}) in the conifold locus $\Delta=0$, only two 
lead to $\kappa \in \IR$. They are the curves in the $(\kappa, B)$ plane with 
$\kappa=\pm 4 \cos\pi B$, which correspond respectively to $a=1$ and $b=1$, 
therefore to $\lambda_1=0$ or $\lambda_2=0$. Hence, the boundary of the ABJM moduli space 
given by $\text{min}(\lambda_1, \lambda_2)=0$ corresponds to
\begin{equation}
\label{clocusb}
 \kappa(B)= \left\{
     \begin{array}{cl}
       -4\cos\pi B\,, &\qquad  B>1/2\\
       4\cos\pi B\,, &\qquad  B<1/2
     \end{array}
   \right.
\end{equation} 
In particular, the symmetric conifold point $z_1=z_2=1/16$ corresponds to 
$B=n \in \IZ$, $\kappa=\pm 4$.
Along the curve (\ref{clocusb}), one of the two gauge groups of the ABJM 
theory is absent, so the theory reduces to topological CS theory. We examine 
this regime in Section~\ref{sec:conifold}.

\end{enumerate}

Given a fixed value of the $B$ field, we can describe the real one-dimensional moduli space of the ABJM theory 
as a real submanifold of the $u$-plane of Seiberg--Witten theory, by using (\ref{umodulus}) in the form
\be
u=-\cos(2 \pi B) + {\kappa^2 \over 8}. 
\ee
Singular points in moduli space become then 
the well-known singularities of SW theory. For example, when $B=1/2$, the moduli space, described by $\kappa \in [0, \infty)$, becomes the region 
$u \in [1, \infty)$. The orbifold point (weakly coupled ABJM theory) maps to the monopole point, while the large radius point 
(strongly coupled ABJM theory) corresponds to the semi-classical region (see \figref{uplanefig}). Notice that the conifold point 
would map to the dyon point of Seiberg--Witten theory, but this does not belong to the moduli space of ABJM theory with $B=1/2$. We can however 
realize it by making an analytic continuation of the 't~Hooft coupling to complex values. The dyon point corresponds then to the point $\kappa^2=-16$, which leads by (\ref{lamkap}) to an 
imaginary value
\be
\lambda= -{2 \ri \text{K} \over \pi^2},
\ee
where $\text{K}$ is Catalan's number. 

As usual, string dualities lead to a full complexification of the moduli space of 't~Hooft parameters. In the case of ABJM theory, the complexified moduli space for the variables 
$\lambda_{1,2}$ is simply the moduli space of the parameters $\beta, \zeta$, which is a $\IZ_2 \times \IZ_2$ covering of the moduli space parametrized by $z_{1,2}$.

\FIGURE{
\includegraphics[height=6cm]{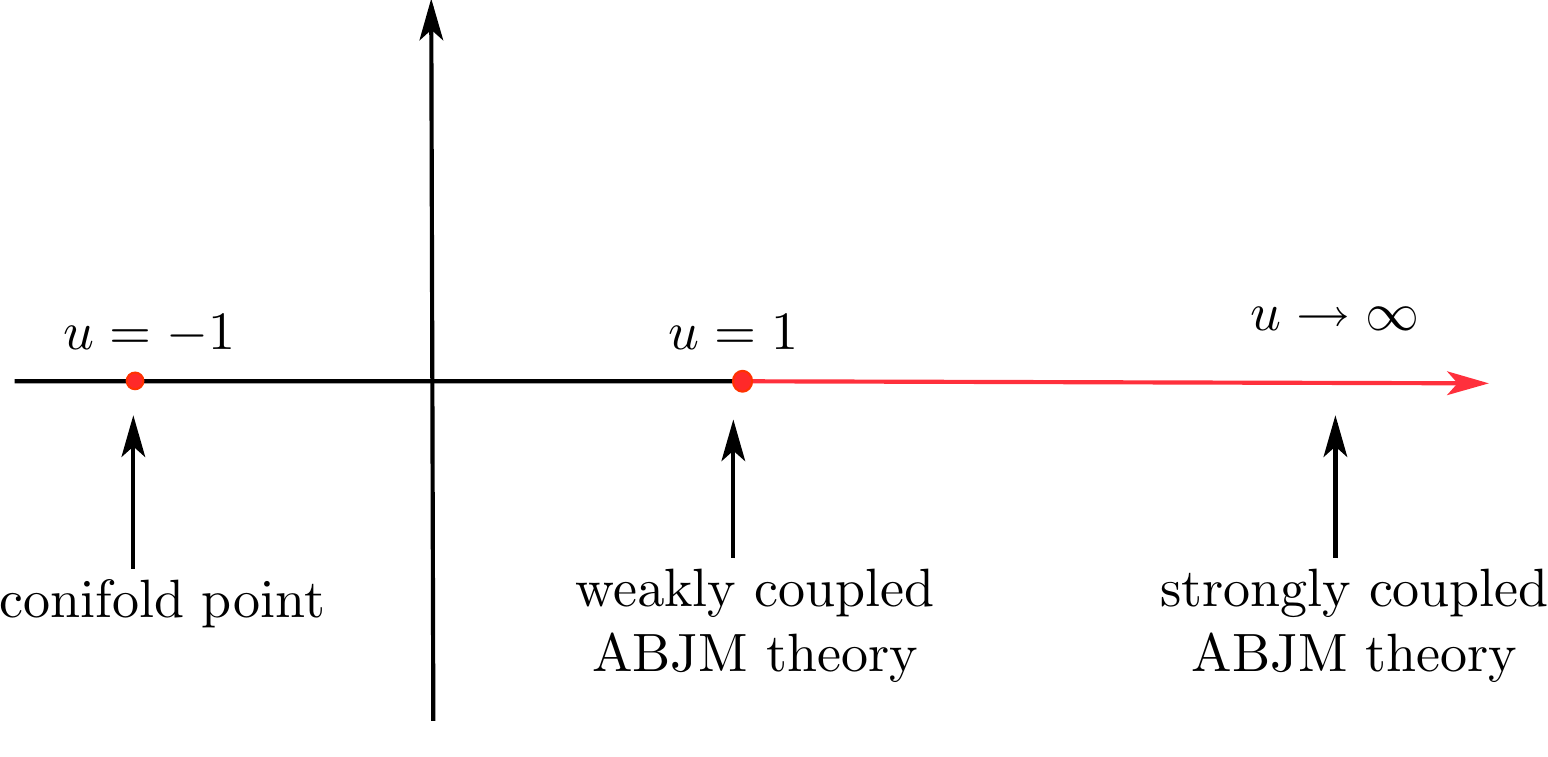} 
\caption{The moduli space of the ABJM theory for $B=1/2$ can be mapped to the line $[1, \infty)$ in the $u$ plane of Seiberg--Witten theory, 
which is here shown in red. 
The monopole point corresponds to the weakly coupled ABJM theory, while the semiclassical limit corresponds to the 
strongly coupled theory.}
\label{uplanefig}
}

\sectiono{Weak coupling}
\label{sec:weak}

In principle, to study the matrix model at weak coupling one does not need the 
sophisticated tools presented in the previous section. One can do perturbative 
calculations directly in the integral expressions (\ref{kapmm}) or (\ref{intdef}) 
for the matrix model. A calculation of the $1/6$ BPS Wilson loop to three loop order 
was indeed done in this way in the original paper \cite{kapustin}.

Still, the explicit expressions for the periods $\sigma_{1,2}$ (\ref{speriods}) and 
their relation to $t_{1,2}$ (\ref{sigma-t}) gives a much more efficient way to obtain perturbative, planar 
expansions. Inverting 
these relations we find the weak coupling expression for $\kappa$ (\ref{zetakappa})
\be
\label{weakkappa}
\ba
\kappa=&\,
-2 \ri (t_1-t_2)
-\frac{\ri}{12} \left(t_1^3+3 t_1^2 t_2-3 t_1 t_2^2-t_2^3\right)
\\&\,
-\frac{\ri}{960}\left(t_1^5+5 t_1^4 t_2-10 t_1^3 t_2^2+10 t_1^2 t_2^3-5 t_1t_2^4-t_2^5\right)
+\CO(t^7).
\ea
\ee
This agrees with the weak coupling expansion of the inverse of the exact mirror map 
(\ref{lamkap}), obtained in \cite{mp}.

Using the dictionary relating the 't~Hooft couplings (\ref{relthoofts}) we immediately get 
the result for the $1/2$ BPS Wilson loop in the planar approximation (\ref{12planar})
\be
\label{12weak}
\ba
\langle W^{1/2}_{\tableau{1}} \rangle
=\re^{\pi\ri B}\frac{\kappa}{2}=
&\,
\re^{\pi\ri(\lambda_1-\lambda_2)}\,2 \pi\ri  (\lambda_1+\lambda_2)
\bigg[1
-\frac{\pi^2}{6} \left(\lambda_1^2-4 \lambda_1\lambda_2+\lambda_2^2\right)
\\&\,\hskip2cm
+\frac{\pi ^4 }{120}\left(\lambda_1^4-6\lambda_1^3\lambda_2-4\lambda_1^2\lambda_2^2
-6\lambda_1\lambda_2^3+\lambda_2^4\right)
+\CO(\lambda^6)\bigg].
\ea
\ee
In this expression we factored out the term $2 \pi\ri  (\lambda_1+\lambda_2)$, which 
depends on the overall normalization of the Wilson loop, as mentioned after (\ref{12planar}). 
There is also the extra phase factor, which appears also at strong coupling and 
can be attributed to framing. 
Note that so far this expansion has not been reproduced directly in the gauge theory, as 
even the two-loop graphs are quite subtle.

For the $1/6$ BPS Wilson loop, using the explicit expression (\ref{16-explicit}) and  
expanding at low orders one finds \cite{mp}
\be
\label{16weak}
\left\langle W^{1/6}_{\tableau{1}} \right\rangle
=\re^{\pi\ri\lambda_1}2\pi\ri\lambda_1
\left(1-\frac{\pi ^2}{6} \lambda_1(\lambda_1-6 \lambda_2)
-\frac{\pi ^3\ri}{2} \lambda_1 \lambda_2^2
+\frac{\pi ^4}{120} \lambda_1 \left(\lambda_1^3-10 \lambda_1^2\lambda_2-20 \lambda_2^3\right)
+\CO(\lambda^5)\right).
\ee
Again the exponent is a framing factor and the factor of $2\pi\ri\lambda_1$ is due to the 
normalization chosen in (\ref{1/6-integral}). This expression agrees with the 2-loop 
calculations in \cite{dp,cw,rey}. Note that the 3-loop analysis in \cite{rey}, 
done for $\lambda_1=\lambda_2$, misses the next term, due to a projection 
which essentially removes all terms at odd orders in perturbation theory.

Next we turn to the free energy. Here we notice that  the period in (\ref{pder}) gives only 
the derivative of the free energy. Indeed, within the formalism of special 
geometry developed above, the planar free energy of the matrix model is only 
determined up to quadratic terms in the 't~Hooft couplings. These have to be 
fixed by direct calculation in the matrix model
\be
F={N_1^2 \over 2} \log \left({2\pi N_1 \over  k}\right) +{N_2^2 \over 2} \log \left({2\pi N_2 \over  k}\right) -{3\over 4} (N_1^2 + N_2^2) -\log(4) N_1 N_2 +\cdots 
\label{Fweak}
\ee
The last term comes from 
the normalization of the $\cosh$ term in (\ref{kapmm}), while the remaining terms are 
just the free energies for two Gaussian matrix models with couplings $\pm 2 \pi \ri/k$. 
Notice that the above free energy has an imaginary piece given by 
\be
{\pi \ri \over 6 k}(N_1-N_2)((N_1-N_2)^2-1).
\label{F-imaginary}
\ee
Using the identification of the periods at weak coupling (\ref{I-weak}) we write down the 
next term in the perturbative expansion
\be
\frac{\pi^2}{72k^2} \left(N_1^4-6 N_1^3 N_2+18 N_1^2 N_2^2
-6 N_1 N_2^3+N_2^4\right).
\ee
It would be interesting to try to reproduce these expressions directly from studying 
perturbative ABJM theory on $\IS^3$.

\sectiono{Strong coupling expansion and the AdS dual}
\label{sec:strong}

We turn now to the strong coupling limit of the matrix model, where we have to find 
the analytic continuation of the 't~Hooft parameters to the strong coupling region, 
as functions of the global parameters of moduli space. 
We will see how the shift of the charges discussed in \cite{bergman, ah} 
emerges naturally from our computation. We will also evaluate the free energy in 
this regime and compare with the classical action of the vacuum AdS dual, deriving in this way the $N^{3/2}$ behavior 
of the degrees of freedom.

\subsection{Analytic continuation and shifted charges}
\label{sec:analytic}

In order to perform the analytic continuation of the 't~Hooft parameters, we 
use the explicit representation of the periods in terms of integrals 
given in (\ref{tperiods}) as well 
as their derivatives (\ref{derperzeta})-(\ref{derperbeta}). 
Let us start by discussing $t_1$. We study its behavior at large $\zeta$ but fixed $\beta$, which is the large radius 
region. We find
\begin{equation}
 \frac{\d t_1}{\d\zeta}=\frac{\ri}{\pi\zeta}\,\log\left(-\frac{\zeta^2}{\beta}\right)+o(\zeta^{-1}), \qquad  
 \frac{\d t_1}{\d\beta}=-\frac{\ri}{2\pi\beta}\,\left(\log(-\zeta^2)+\pi \ri\right)+o(1),
\end{equation} 
and this gives the leading behavior
\begin{equation}
t_1=-\frac{\ri}{2\pi}\left(\log(-\zeta^2)+\pi\ri\right)\log\frac{\beta}{\zeta}+\cdots
\label{asympt_zeta}
\end{equation} 
In the physical theory $t_1$ should be imaginary and $\beta$ a phase. By examining 
(\ref{asympt_zeta}), this implies that $\kappa$ is real. From (\ref{z1z2-Bkappa}) we 
then see that $z_1=\bar z_2$ and henceforth we label it $z_1=z$. 

We know also that $t_1$ must be a linear combination of the periods 
at large radius. Using that $z_1=1/\zeta^2$ and $z_2=(\beta/\zeta)^2$, and comparing 
(\ref{asympt_zeta}) to the behavior of the periods (\ref{mirrormapone}) and (\ref{dperiods}), we find 
\begin{equation}
\ba
 t_1&=\frac{\ri}{2\pi}(F_1+F_2)-\frac{1}{2}T_2-\frac{\pi \ri}{6},\\
 t_2&=-\frac{\ri}{2\pi}(F_1+F_2)+\frac{1}{2}T_1+\frac{\pi \ri}{6}.
 \ea
\label{tFFT}
\end{equation} 
The constants $\pm\pi \ri/6$ cannot be fixed by using the above information, but 
they can be fixed by specializing to the ABJM slice $z_1=z_2$, as we will see in a moment.

A simple calculation leads to the following explicit expression 
\be
\label{lam1}
\lambda_{1}(\kappa, B)={1\over 2}\left( B^2 -{1\over 4}\right) +\frac{1}{24}+ {\log^2\kappa \over 2 \pi^2}-{\log\kappa \over 2 \pi^2} \omega^{(1)}\left( z, \bar z \right) 
 +{1\over 16 \pi^2} \left( \omega^{(2)}_1+\omega^{(2)}_2\right)\left(z, \bar z \right).
\ee
This expansion is valid in the region $\kappa \rightarrow +\infty$. Notice that it is manifestly real when $\kappa$ is real and positive.

As a check of the above expression, we can particularize to the ABJM slice 
$\lambda_1=\lambda_2=\lambda$, ($B=1/2$), which corresponds in the 
gauge theory, to having identical gauge groups in the two nodes of the quiver, {\em i.e.}, 
$N_1=N_2$. The mirror map for this case was obtained in \cite{mp} as
\be
\label{lamkap}
 \lambda\left(\kappa,B={1\over 2}\right)={\kappa \over 8 \pi}   
 {~}_3F_2\left(\frac{1}{2},\frac{1}{2},\frac{1}{2};1,\frac{3}{2};-\frac{\kappa^2}{16}\right).
\ee
The strong coupling expansion of this expression at $\kappa\gg 1$ is
\be
 \lambda\left(\kappa,B={1\over 2}\right) 
 ={\log ^2\kappa\over 2 \pi^2}+\frac{1}{24}+\CO( \kappa^{-2}),
\ee
in agreement with (\ref{lam1}). This also fixes the constants in (\ref{tFFT}).

As in \cite{mp}, the observables of the model are naturally functions of $\zeta$, $\beta$ 
(alternatively $\kappa$, $B$), and we have to re-express them in terms of 
$\lambda_{1,2}$. Equation (\ref{lam1}) 
shows that the natural variable at strong coupling is not $\lambda_1$, but rather 
\be
\label{hlam}
\hat \lambda = \lambda_1-{1\over 2} \left( B^2 -{1\over 4}\right) -\frac{1}{24}
=\frac{1}{2}(\lambda_1+\lambda_2)-\frac{1}{2}(\lambda_1-\lambda_2)^2-\frac{1}{24}. 
\ee
In particular, it is only when expressed in terms of this variable that $\kappa$ is a periodic function of $\hat \lambda, B$. 

Remarkably, the above shift is precisely the one 
found in \cite{ah}. In the type IIA realization of the ABJ theory $U(M_2)_k \times U(M_2+M_4)_{-k}$, where $M_2$ corresponds to the number of D2 branes and 
$M_4$ to the number of D4 branes, the Maxwell charge of the D2 branes is not $M_2$, but rather 
\be
Q_2=M_2 -{k\over 2} \left( B^2 -{1\over 4}\right) -{1\over  24}\left( k -{1\over k}\right),
\ee
where
\be
B=-{M_4 \over k} +{1\over 2}. 
\ee
After dividing by $k$ and taking the large $k$ limit, we recover (\ref{hlam}) with 
\be
\hat \lambda={Q_2 \over k}. 
\ee

The relation between $\hat \lambda$ and $\kappa$ can be inverted at strong coupling, generalizing \cite{mp} to $B\neq\frac{1}{2}$, and it is of the form 
\be
\label{fullkappa}
\kappa(\hat \lambda, B) =\re^{\pi {\sqrt{2\hat \lambda}}} \left( 1+ \sum_{\ell\ge 1} c_\ell \left({1\over \pi {\sqrt{2 \hat \lambda}}} , \beta \right) \re^{-2 \ell  \pi {\sqrt{2\hat \lambda}}}  \right)
\ee
where 
\be
c_\ell(x, \beta) =\sum_{k=0}^{2\ell-1} c_k^{(\ell)} (\beta) x^k.
\ee
The coefficients $c_k^{(\ell)}(\beta)$ are Laurent polynomials in $\beta, \beta^{-1}$, of degree $\ell$, and symmetric 
under the exchange $\beta \leftrightarrow \beta^{-1}$. In other words, 
they can be written as polynomials in $\cos(2 \pi m B)$, so they are periodic in $B$, with period $1$. We find, for example, 
\be
\ba
c_1(x, \beta)&=-\left( \beta + \beta^{-1} \right) \left(1 -{x\over 2 }\right),\\
c_2(x, \beta)&=3+ {x \over 8} \left( 3 \beta^2 -8 + 3 \beta^{-2}\right)- {3 x^2  \over 8} \left( \beta + \beta^{-1} \right)^2 -{x^3 \over 8} \left( \beta + \beta^{-1} \right)^2. 
\ea
\ee
The fact that $c_\ell(x, \beta)$ are polynomials in $x$ of degree $2\ell-1$, rather than power series, comes out from an explicit calculation of the first few cases, and we have not 
established it. 

From the explicit expression (\ref{lam1}) we can implement the symmetries of the model as a function of $\kappa$ and $B$ (or equivalently, $z_1$ and $z_2$). 
For example, the transformation 
\be
N_1 \rightarrow 2 N_1 +k -N_2, \qquad N_2 \rightarrow N_1
\label{monodromy}
\ee
simply corresponds to periodicity in the $B$ field
\be
B \rightarrow B+1
\ee
while $\kappa$ remains unchanged. From the point of view of the $z_{1,2}$ variables, this is simply a monodromy transformation $z_{1,2} \rightarrow \re^{\mp 2 \pi \ri} z_{1,2}$. 
Notice that not all the values of $\kappa$ lead to admissible values of $\lambda_{1,2}$, since $\text{min}(\lambda_1, \lambda_2) \ge 0$. This means that the boundary of moduli space 
is the conifold locus (\ref{clocusb}).

\subsection{Wilson loops at strong coupling and semi--classical strings}
\label{sec:WL-strong}

As an application of the explicit expression for $\kappa$ (\ref{fullkappa}), 
we can use (\ref{zetakappa}) to immediately obtain 
the VEV of the $1/2$ BPS Wilson loop (\ref{12planar}) at strong coupling
\be
\label{12strong}
\langle W^{1/2}_{\tableau{1}} \rangle_{g=0}
=\frac{1}{2}\,\re^{\pi \ri B} \kappa(\hat \lambda, B).
\ee
Note that this is a real function of $\hat \lambda, B$, up to the overall phase involving 
the $B$ field. This is the same phase that appears also in the weak-coupling result 
(\ref{12weak}) and arises also in field theory calculations as a framing-dependant 
term \cite{witten,gmm,al}. The matrix model always gives the answer for framing=1. 

The result for the $1/6$ BPS Wilson loop is, as usual more complicated, but can 
still be written in a power series expansion at strong coupling. We quote only the 
leading strong coupling result for $\lambda_1=\lambda_2$ \cite{mp}
\be
\langle W^{1/6}_{\tableau{1}} \rangle_{g=0}
\approx-\frac{\sqrt{2\lambda}}{4}\,\re^{\pi \ri \lambda_1}\,\re^{\pi\sqrt{2\lambda}}.
\ee

We would like to comment about the normalization of the operators. As mentioned 
after (\ref{12planar}), the normalization chosen there is such that the trace of 
the identity in the fundamental of $U(N_1)$ gives $t_1=2\pi\ri N_1/k$ and for 
the fundamental of $U(N_1|N_2)$ (with a minus sign as in (\ref{U}), 
it gives $t_1-t_2=2\pi\ri(N_1+N_2)/k$. In 
CS theory these normalizations are quite common, but they may be not the most 
natural ones in the ABJM theory.

An alternative normalization is to divide by this term, such that at weak coupling 
the expansion of the Wilson loop will be $\langle W\rangle\sim 1+\cdots$. This is 
the normalization chosen in \cite{mp}, and hence the slight differences in the 
preceding equations from that reference. Note, though, that with such a normalization, 
one would have to divide the doubly-wound $1/2$ BPS Wilson loop in the fundamental 
representation by the super-trace of the identity, which is $2\pi\ri(N_1-N_2)/k$ and 
is singular for $N_1=N_2$.

There should be a natural choice of normalization that would reproduce the correct 
normalization fo the one-loop partition function of the classical string in 
AdS$_4\times\IC\IP^3$. To this day, though, a fully satisfactory calculation 
for the analog string in AdS$_5\times\IS^5$ giving the factor of $\lambda^{-3/4}$ 
derived from the the Gaussian matrix model does not exist. One argument, based 
on world-sheet arguments was given in \cite{dg}, but it is not clear why this argument 
would be modified for ABJM theory. Direct calculations of the determinant 
\cite{dgt,kt} were not conclusive. A possible trick to derive it was proposed in 
\cite{drukker} by considering a $1/4$ BPS generalization of the circular Wilson 
loop, where three zero modes of the the Wilson loop of \cite{zarembo} are 
explicitly broken and the integral over them gives this factor. It would be interesting 
to construct such generalization to the Wilson loop of \cite{dt} and see if 
a similar argument can be derived from that.

Regardless of the overall normalization, one can compare those of the $1/2$ BPS loop 
and the $1/4$ BPS loop. Ignoring numerical constants and the framing factor, 
the ratio is
\be
\frac{\langle W^{1/6}_{\tableau{1}} \rangle_{g=0}}
{\langle W^{1/2}_{\tableau{1}} \rangle_{g=0}}
\approx\sqrt\lambda,
\ee
which is proportional to the volume of a $\IC\IP^1$ inside $\IC\IP^3$. Indeed, it 
was argued in \cite{dp,rey} that the string description of the $1/6$ BPS Wilson loop 
should be in terms of a string smeared over such a cycle.

\subsection{The planar free energy and a derivation of the $N^{3/2}$ behaviour}
\label{sec:free}

In this section we study the free energy at strong coupling. We derive the $N^{3/2}$ 
behavior characteristic of M2 branes \cite{klebanov}, and we match the 
exact coefficient with a gravity calculation in type IIA superstring on $\text{AdS}_4\times \IC\IP^3$.

The free energy of the matrix model has a large $N$ expansion of the form
\be
F=\log Z
=\sum_{g=0}^{\infty} g_s^{2g-2} F_g(\lambda_1, \lambda_2).
\label{FE-expand}
\ee
This is the way the genus expansion is typically expressed in topological string theory. 
To compare with the gauge theory and the AdS dual one may choose to rewrite this 
series as an expansion in powers of $1/N$ by absorbing factors of $\lambda$ into $F_g$. 

As mentioned in Section~\ref{sec:weak}, the formalism of special geometry 
determines the planar free energy only up to quadratic terms 
in the 't~Hooft couplings, and these have to be fixed from the explicit 
weak coupling calculation in the matrix model (\ref{Fweak}).

Let us now consider the derivative of the genus zero free energy (\ref{pder}), 
and study its analytic continuation to strong coupling as we have done 
for $t_i$ at the top of Section~\ref{sec:analytic}. Expanding (\ref{I-derivs}) for 
large $\kappa$ we find
\be
\frac{\partial\CI}{\partial\zeta}=-\frac{\pi\ri}{\zeta}+\CO(\zeta^{-2}),\qquad
\frac{\partial\CI}{\partial\beta}=\CO(\zeta^{-1}),
\ee
so
\be
\CI=-\pi\ri\log\zeta+\CO(\zeta^0)=-\pi\ri \log \kappa +\pi^2 B + \CO(\kappa^0,B^0), \qquad \kappa \to \infty, 
\ee
From this leading large $\kappa$ behavior we have that in the ABJM slice 
\be
\frac{\partial F_0}{\partial\lambda}\approx2\pi^3\sqrt{2\lambda}, 
\ee
which can be integrated 
to give the leading term in (\ref{prepotf}) and the match with the supergravity calculation 
presented below.

But to get the full series of corrections we should proceed more carefully. 
We know that the result of the continuation should be a linear combination of periods, 
and comparing to (\ref{mirrormapone}) we see that we can express the period as
\be
\label{freefirst}
\CI+\frac{\pi\ri t}{2}=\frac{ \partial F_0}{\partial t_1}-\frac{ \partial F_0}{\partial t_2} 
= -{\pi \ri \over 4} \left( T_1+T_2 +2\pi \ri \right). 
\ee
The constant term can be fixed by looking at the solution on the ABJM slice $N_1=N_2$, 
which can be obtained as follows. Since on the slice we effectively have a one-parameter model, there is only 
one Yukawa coupling, which we can integrate to obtain $F_0$. From (\ref{yuka}) we easily obtain
\be
\label{yukawa}
\partial_{\lambda}^3 F_0(\lambda)={1\over 4} C_{\lambda \lambda\lambda}\Bigl|_{\lambda_1=-\lambda_2}=
-{128 \pi^6 \over \kappa (\kappa^2+16)}{1\over K\left({\ri \kappa \over 4}\right)^3}
\ee
where the factor of $4$ is introduced to match the normalization of the matrix model, and we used that 
\be
  {\rd \lambda \over \rd \kappa}={1\over 4 \pi^2} K\left({\ri \kappa \over 4}\right). 
\ee
Integrating once, we find
\be
\label{seconder}
\partial_{\lambda}^2 F_0(\lambda) =4 \pi^3  {K'\left({\ri \kappa \over 4}\right)\over K \left({\ri \kappa \over 4}\right)} + a_1, 
\ee
where $a_1$ is an integration constant and we have used the Legendre relation
\be
E'K + EK'-K K'={\pi \over 2}. 
\ee
A further integration leads to the following expression in terms of a Meijer function
\be
\partial_\lambda F_0 (\lambda)={\kappa\over 4} G^{2,3}_{3,3} \left( \begin{array}{ccc} {1\over 2}, & {1\over 2},& {1\over 2} \\ 0, & 0,&-{1\over 2} \end{array} \biggl| -{\kappa^2\over 16}\right) + a_1 \lambda + a_2.
\ee
Comparison with the matrix model free energy at weak coupling (\ref{Fweak})
fixes $a_1=4 \pi^3 \ri$, $a_2=0$, so we can write
\be
\label{comf}
 \partial_\lambda F_0 (\lambda)={\kappa \over 4} G^{2,3}_{3,3} \left( \begin{array}{ccc} {1\over 2}, & {1\over 2},& {1\over 2} \\ 0, & 0,&-{1\over 2} \end{array} \biggl| -{\kappa^2\over 16}\right)+{ \pi^2 \ri \kappa \over 2} 
  {~}_3F_2\left(\frac{1}{2},\frac{1}{2},\frac{1}{2};1,\frac{3}{2};-\frac{\kappa^2.
   }{16}\right).
\ee
If we integrate this expression with the following choice of integration constant, 
\be
\label{correctweak}
  F_0(\lambda)=\int_0^{\lambda} \rd \lambda' \,  \partial_{\lambda'} F_0 (\lambda')
\ee
we obtain the correct weak coupling expansion.  

We can now analytically continue the r.h.s. of (\ref{comf}) to $\kappa=\infty$, 
and we obtain
\be
\label{slice}
\partial_\lambda F_0 (\lambda)=2\pi^2 \log \kappa +{4 \pi^2 \over \kappa^2} \, {}_4 F_3 \left( 1, 1, {3\over 2}, {3\over 2}; 2,2,2; -{16 \over \kappa^2} \right)
\ee
This agrees with (\ref{freefirst}) on the ABJM slice. To see this, one notices that
\be
\ba
\omega^{(1)}(z, z)&=2 \sum_{n=1}^{\infty} \sum_{k+l=n} {(2k + 2l -1)! \over (k!)^2 (l!)^2 } z^n=
 2 \sum_{n=1}^{\infty} \frac{4^n (2 n-1)! \Gamma \left(n+\frac{1}{2}\right)}{\sqrt{\pi } \Gamma (n+1)^3} z^n\\
 &=4 z \, {}_4 F_3 \left( 1, 1, {3\over 2}, {3\over 2}; 2,2,2; 16 z \right)
 \ea
\ee
is precisely the generalized hypergeometric function appearing in (\ref{slice}).

We are now ready to discuss the calculation of the planar free energy at strong coupling. We have, 
\be
\label{prepre}
\partial_{\hat \lambda} F_0 (\lambda_1, \lambda_2) =2\pi^2 \log\kappa -\pi^2 \omega^{(1)}(z, \bar z).
\ee
After plugging the value of $\kappa$ in terms of $\hat \lambda$ given by the series expansion (\ref{fullkappa}), and integrating w.r.t. $\hat \lambda$, we obtain 
\be
\label{prepotf}
F_0 (\hat \lambda, B)= {4\pi^3 \sqrt{2}  \over 3} \hat \lambda^{3/2} +{\zeta(3) \over 2}
+\sum_{\ell\ge1}  \re^{- 2\pi \ell  {\sqrt{2\hat \lambda}}} f_{\ell}\left({1\over \pi {\sqrt{2 \hat \lambda}}},\beta \right)-\frac{2\pi^3 \ri}{3}\left(B-\frac{1}{2}\right)^3,
\ee
where $f_{\ell}(x)$ is a polynomial in $x$ of the form 
\be 
f_{\ell}(x,\beta)=\sum_{k=0}^{2\ell-3} f_k^{(\ell)} (\beta) x^k, \qquad \ell\ge 2. 
\ee
The coefficients $f_k^{(\ell)}(\beta) $ are Laurent polynomials in $\beta$ of degree $\ell$, and symmetric under the exchange 
$\beta \leftrightarrow \beta^{-1}$. We have, 
for the very first cases, 
\be
\ba
f_1(x,\beta)&=- {1\over 2}\left( \beta + \beta^{-1} \right),\\
f_2(x,\beta)&=\frac{1}{16} \left(\beta ^2+16+\beta^{-2}\right)+ \frac{x}{4 }\left(\beta +\beta^{-1}\right)^2.
\ea
\ee
In going from (\ref{prepre}) to (\ref{prepotf}) an integration constant $\zeta(3)/2$ appears. Its presence can be checked by comparing (\ref{prepotf}) with a numerical calculation of the integral (\ref{correctweak}) at intermediate coupling\footnote{This integration constant was incorrectly set to zero in a previous version of the paper. It was determined numerically in \cite{hanada}.}. This constant is nothing but the well-known constant map contribution to the prepotential, first found in \cite{cdgp}. 

The free energy in the planar approximation is given by rescaling (\ref{prepotf}) by the 
string coupling $F= g_s^{-2}F_0+\CO(g_s^0)$. This expression  displays many interesting 
features.  First, note that on the ABJM slice $N_1=N_2$ the leading term 
\be
\label{ffree}
-{\pi\sqrt{2} \over 3} k^2  \hat\lambda^{3/2}
\ee
displays the ``anomalous" scaling 
$N^{3/2}$ in the number of degrees of freedom for a theory of M2 branes, 
as was first derived from a supergravity calculation in \cite{klebanov}. 
The above calculation is a first principles derivation 
of this behaviour at strong coupling in the gauge theory. Usually, this behaviour is 
associated to the thermal free energy on $\IR^3$, while (\ref{ffree}) gives rather the free 
energy of the ABJM theory on $\IS^3$ at strong coupling. However, a supergravity 
calculation of this free energy also leads to the $N^{3/2}$ behavior. We will 
show this now, and in particular we will match the numerical coefficient 
in (\ref{ffree}).\footnote{We would like to thank Diego Hofman 
for very useful remarks on this calculation.}

\subsection{Calculation of the free energy in the gravity dual} 

Consider type IIA theory on AdS$_4 \times \IC\IP^3$, and let us reduce it to the AdS$_{4}$ factor as in for example \cite{by}. 
The (Euclidean) AdS metric appropriate for a boundary theory on $\IS^3$ is
\be
\rd s^2= \rd\rho^2 + \sinh^2\rho\,  \rd \Omega^2,
\ee
where $\rd \Omega^2$ is the metric on an $\IS^3$ of unit radius. In this coordinate system, the boundary is at $\rho \to \infty$. The free energy of the 
boundary CFT on $\IS^3$ should be given, in the supergravity approximation, by minus the Euclidean gravitational action of the AdS$_{n+1}$ space $-I_{\text{AdS}_{n+1}}$. This action 
is given by a bulk term, a surface term, and a counterterm at the boundary \cite{bk,ejm}
\be
I_{\text{AdS}_{n+1}}=I_\text{bulk}+I_\text{surf}+ I_\text{ct}, 
\ee
with 
\be
\label{difints}
\ba
I_\text{bulk}&=-{1\over 16 \pi G_N} \int_X \rd^{n+1}x\, {\sqrt{g}} \left( R-2 \Lambda\right),\\
I_\text{surf}&=-{1\over 8 \pi G_N} \int_{\partial X} \rd^n x\, {\sqrt{h}} K, \\
I_\text{ct}&={1\over 8 \pi G_N} \int_{\partial X} \rd^n x\, {\sqrt{h}} \left[ n-1 + {1\over 2(n-2)} \CR +\cdots\right]. 
\ea
\ee
In these equations, $G_N$ is Newton's constant, and $R$, $K$ and $\CR$ are the scalar curvature of the bulk, the extrinsic curvature of the boundary $\partial X$, and the 
scalar curvature of the induced metric $h$ on $\partial X$, respectively. The counterterm action includes higher order corrections which are not relevant for the 
case of AdS$_4$ and will not be considered here \cite{ejm}. As our boundary $\partial X$, we will take the hypersurface $\rho=\rho_0$, and at the end of the calculation 
we must take $\rho_0 \to \infty$. The counterterms guarantee that the resulting action will be finite. 

The bulk action is easy to evaluate and gives
\be
I_\text{bulk}(\rho_0)={3\over 8 \pi G_N}  \text{vol} (\text{AdS}_4; \rho_0)
\ee
where 
\be
\label{volreg}
\text{vol} (\text{AdS}_4; \rho_0)=\text{vol} (\IS^3)  \int_0^{\rho_0} \rd \rho \, (\sinh \rho)^3 =
2 \pi^2  \left[ {1\over 12} \cosh(3 \rho_0) -{3\over 4} \cosh(\rho_0) +  {2\over 3}\right].
\ee
It is easy to see that the surface term and the counterterms remove the divergences as $\rho_0 \to \infty$, leaving only the term 
$ 4 \pi^2/3$ in (\ref{volreg}), and we find \cite{ejm}
\be
\lim_{\rho_0 \to \infty} I_{\text{AdS}_4}(\rho_0)= {\pi \over 2 G_N}. 
\ee
If we now use the dictionary relating Newton's constant to the gauge theory data, 
\be
{1 \over G_N}={2  \sqrt{2} \over 3} k^2 \hat \lambda^{3/2},
\ee
we find exactly the leading term in (\ref{prepotf})! Of course, in order to obtain this result we have used the regularization provided by the counterterm integral in (\ref{difints}), 
and one could suspect that the matching depends very much on this regularization. However, this 
counterterm has been tested (or fixed) in an independent way in the calculations of \cite{bk,ejm}. In particular, for $n=4$ it leads to the matching of the 
Casimir energy of $\CN=4$ SYM on $\IR \times \IS^3$, and for $n=3$ it reproduces the standard mass of an AdS$_4$--Schwarzschild black hole \cite{bk}. Therefore,  
the above calculation provides a genuine test of the AdS$_4$/CFT$_3$ correspondence. 

\FIGURE{
\includegraphics[height=4cm]{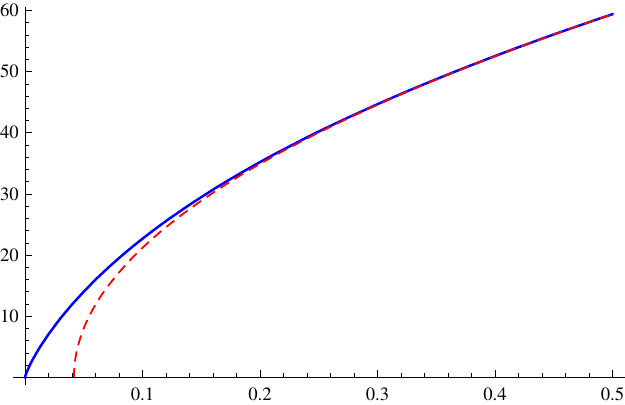} \qquad 
\includegraphics[height=4cm]{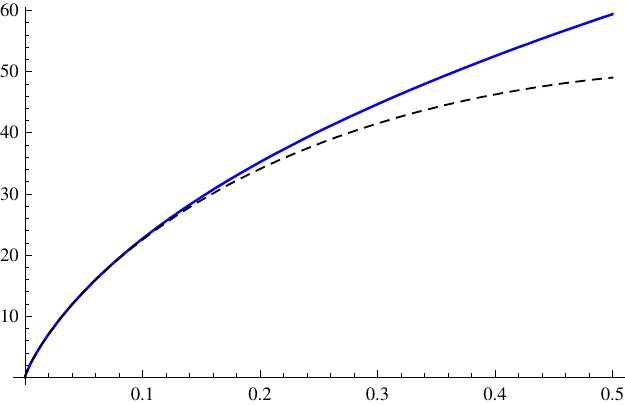}
\caption{Comparison of the exact result for $\partial_\lambda F_0(\lambda)$ given in (\ref{comf}), plotted as a solid blue line, and the weakly coupled and 
strongly coupled results. In the figure on the left, the red dashed line is the supergravity result (\ref{sugracomp}), while in the figure on the right, the 
black dashed line is the Gaussian result (\ref{Gcomp}).}
\label{prepotfig}
}

In \figref{prepotfig} we show the exact result for the planar limit of $\partial_\lambda F_0(\lambda)$ in the case $N_1=N_2$, as a function of $\lambda=N/k$, and we compare it 
to the behavior of the supergravity prediction 
\be
\label{sugracomp}
\partial_\lambda F_0(\lambda) \approx 2 \pi^{3} {\sqrt{2 (\lambda-1/24)}}, \qquad \lambda\to \infty.
\ee
We see that the strong coupling behavior gets triggered for values of the coupling $\lambda \approx 0.2$. For $\lambda \to 0$, the behavior of the prepotential is 
dominated by the Gaussian, weakly coupled result (\ref{Fweak})
\be
\label{Gcomp}
 \partial_\lambda F_0(\lambda)
 \approx -8\pi^2 \lambda\left( \log\left( {\pi \lambda \over 2} \right)-1\right),
 \qquad \lambda\to 0. 
\ee

A second aspect to notice is that the supergravity result (\ref{prepotf}) has corrections which are exponentially suppressed. The exponential is of the form 
\be
\re^{-\ell A(\IC \IP^1)}
\ee
where
\be
A(\IC \IP^1) =2 \pi {\sqrt{2 \hat \lambda}} 
\ee
is the area of the $\IC\IP^1$ two-cycle in $\IC\IP^3$. Also, notice that each of these exponential corrections multiplies (at each order in $\hat \lambda^{-1/2}$) the polynomial $f_k^{(\ell)}(\beta)$ 
in $\beta, \beta^{-1}$. Therefore, we have contributions schematically of the form 
\be
\sum_{n_+ +n_-=\ell} c_{n_+,n_-}\re^{-n_+ (A(\IC \IP^1) + 2 \pi \ri B) -n_- ( A(\IC \IP^1)- 2 \pi \ri B)}
\ee
This is precisely what one should expect for a gas of $n_+$ instantons and $n_-$ anti--instantons in a $\sigma$ model on $\IC\IP^3$, where the (anti)instantons wrap the $\IC\IP^1$ cycle. 
Notice that this kind of corrections are made possible by the non-trivial topology of two cycles in $\IC\IP^3$, {\em i.e.}, by the fact that $b_2(\IC\IP^3)=1$, and as such they are absent in AdS$_5 \times \IS^5$. Some aspects of these string instantons have been studied in \cite{sorokin}. It would be interesting to test in detail the possible connection between these string instantons and the 
exponentially suppressed corrections to the planar free energy.

These instanton corrections are also present in the Wilson loop result (\ref{12strong}), 
again with an infinite series of corrections. This can be compared with the case 
of $\CN=4$ SYM in 4d, where the asymptotic large coupling expansion of the 
Gaussian matrix model (\ref{N=4wl}) has a single instanton correction which 
can be explicitly identified with a second saddle point solution 
in AdS$_5\times S^5$ \cite{drukker,gpr}.

Finally, we note that when $N_1 \not= N_2$, the planar free energy (\ref{prepotf}) includes 
an imaginary term proportional to $(B-1/2)^3$, which is derived by the weak coupling 
calculation (\ref{F-imaginary}). In CS 
theory such a term is related to framing \cite{witten}. It would be very interesting to derive this phase in type IIA string theory.

\sectiono{Conifold expansion}
\label{sec:conifold}

The expansion around the conifold locus corresponds to a region in the moduli space of the ABJM model where one of the gauge groups has finite coupling, while 
the other one is weakly coupled. In the lens space matrix model this corresponds to one 't~Hooft parameter being small, and the other of order $1$. In this section 
we will study this regime from three different points of view: the exact planar solution in terms of periods and Picard--Fuchs equations, the matrix model, and the 
gauge theory.

\subsection{Expansion from the exact planar solution}
We can use the exact planar solution to calculate various physical quantities near the conifold locus. For concreteness, we will expand around $t_2=0$ but with $t_1$ arbitrary. The first ingredient we need is an expansion of the global coordinates of moduli space. It turns out that the most convenient method 
is based on the expressions for the periods (\ref{tperiods}). The locus where $t_2=0$ is the line
\be
\zeta=2\beta-2, 
\ee
where the cut $(-b, -1/b)$ collapses to the point $Z=-1$. The derivative of $t_2$ w.r.t. $\zeta$ can then be computed in terms of residues at this point by 
expanding the expression in (\ref{derperzeta}):
\be
-\frac{\d t_2}{\d\zeta}=
\sum_{k\geqslant 0}\frac{1}{4\pi \ri}\oint\limits_{-1}\rd Z\,\frac{H_k(Z,\beta)\,(\zeta-2
\beta+2)^k}{(Z+1)^{2k+1}},
\end{equation} 
where $H_k(Z,\beta)$ are regular at $Z=-1$. This gives a series for $t_2$ in powers of $\zeta-2\beta+2$, 
\be
-t_2=\frac{1}{4\sqrt{\beta }}(\zeta-2\beta+2)
-\frac{1-\beta }{128\, \beta ^{3/2}}(\zeta-2\beta+2)^2
+\frac{9-2 \beta +9 \beta ^2}{12288\, \beta ^{5/2}}(\zeta-2\beta+2)^3
+\CO( (\zeta-2\beta+2)^4)
\ee
which can be easily inverted to
\be
  \zeta=2\beta-2-4 \sqrt{\beta }\, t_2+\frac{1}{2} (1-\beta )\, t_2^2
  +\frac{3+10 \beta+3 \beta ^2}{48\sqrt{\beta }} \,t_2^3 +\CO(t_2^4). 
\ee
As a nice application of this expansion, we can compute the VEV of the $1/2$ BPS Wilson loop around the conifold point, which is given in (\ref{12planar}). 
Using the dictionary (\ref{betalambda}), (\ref{zetakappa}), we find
\be
\label{12coni}
\ba
\re^{-\pi \ri B} \langle W^{1/2}_{\tableau{1}} \rangle_{g=0} 
&=\,2 \sin(\pi  \lambda _1)
+2 \pi \lambda _2   \left(2-\cos (\pi  \lambda _1)\right)
+\pi ^2 \lambda _2^2 \sin (\pi  \lambda _1)\\
&\,+\frac{1}{3} \pi^3 \lambda _2^3 \left(1-5 \cos(\pi  \lambda _1)+3 \cos^2(\pi  \lambda _1)\right) +\CO(\lambda_2^4). 
\ea
\ee
As $\lambda_2 \to 0$, we recover the result for a Wilson loop VEV in $U(N_1)$ CS theory. In the conifold expansion we are then regarding the ABJM theory as a perturbation of $U(N_1)$ CS theory at strong coupling. 

The above result can be also obtained by solving the Picard--Fuchs equation around a point in the conifold locus. Let us choose for example the symmetric conifold point (\ref{symconi}), 
with $B=1$ and $\kappa=4$. This corresponds to the point in the conifold locus with 
\be
\lambda_1={1\over 2} , \qquad \lambda_2=0. 
\ee
The appropriate global 
coordinates near this point are (\ref{conglobal}). We find that $\lambda_2$ is a period solving the PF system (\ref{pfconi}) and with leading behavior
\be
 \lambda_2=-\frac{1}{4\pi}\left(y_2+y_1/2\right)+\mathcal{O}(y^2). 
\ee
One finds the expansion 
\be
\ba
\lambda_2=&\,\frac{\pi}{4}(B-1)^2-\frac{5\pi^3}{96}(B-1)^4
+\left(\frac{1}{8 \pi }-\frac{\pi}{32}(B-1)^2+\frac{43 \pi^3}{1536}(B-1)^4\right) (\kappa -4) 
\\
&\,+\left(-\frac{1}{128 \pi }+\frac{9\pi}{1024}(B-1)^2-\frac{99\pi^3}{8192}(B-1)^4\right) (\kappa -4)^2
+ \CO\big((B-1)^6\big)+\CO\big((\kappa-4)^3\big),
\ea
\ee
which is inverted to 
\be
\ba
 \kappa=&\,4-2 \pi ^2 \left(\lambda _1-\frac{1}{2}\right)^2
 +\frac{\pi ^4}{6} \left(\lambda _1-\frac{1}{2}\right)^4
 +\pi \lambda_2\left(8 +4 \pi \left(\lambda _1-\frac{1}{2}\right)-\frac{2 \pi^3}{3} \left(\lambda _1-\frac{1}{2}\right)^3\right) 
 \\&\,+\CO\big(\lambda_2^2\big)+\CO\big((\lambda_1-1/2)^5\big).
 \ea
\ee
This is indeed the expansion around $\lambda_1=1/2$ of (twice) the series in the r.h.s. of (\ref{12coni}). 

Once we know the expansion of the global coordinates, we can consider other quantities in the model, like the genus $g$ free energies. 
The conifold 
expansion of $F_g(t_1, t_2)$ has the form
\be
\label{fgconex}
F_g(\lambda_1, \lambda_2) =F_g^{\rm G}  (\lambda_2) +\sum_{n\ge 0} F_g^{(n)}(\lambda_1) \lambda_2^n, 
\ee
where $F_g^{\rm G}(\lambda_2)$ is the free energy of the $U(N_2)$ Gaussian matrix model, and each coefficient $F_g^{(n)}(\lambda_1)$ can be obtained 
as an exact function of $\lambda_1$. 
Of course, 
\be
F_g^{(0)}(\lambda_1) =F_g^{\IS^3}(\lambda_1)
\ee
is the genus $g$ free energy of the CS theory on $\IS^3$. When $g=0$, the expansion (\ref{fgconex}) can be computed from the exact planar 
solution in various ways. One can for example use the Yukawa couplings (\ref{yuka}) expanded around the conifold locus in order to compute the third derivatives of $F_0$, 
or use the modularity properties of the solution discussed in \cite{abk,hkr}. In any case, for the first few functions one finds the following results:
\be
\label{conlist}
\ba
F_0^{(1)}(\lambda_1)&=2\pi\ri\left(\pi^2\lambda_1^2
+2\text{Li}_2\big(-e^{\pi\ri\lambda_1}\big)
-2\text{Li}_2\big(-e^{-\pi\ri\lambda_1}\big)\right),\\
F_0^{(2)}(\lambda_1)  &=-2 \pi^3 \ri \lambda_1 +8 \pi ^2 \log \left(\cos \left(\frac{\pi  \lambda _1}{2}\right)\right),\\
F_0^{(3)}(\lambda_1)  &={2 \pi^3 \ri  \over 3} +\frac{\pi^3}{3}\big(3 \cos(\pi  \lambda _1)-5\big) \tan \left(\frac{\pi  \lambda _1}{2}\right).
  \ea
  \ee

\subsection{Conifold expansion from the matrix model}

It is easy to implement the conifold expansion directly in the lens space matrix model. To do that, 
we notice that it can be written as two interacting Chern--Simons matrix models on $\IS^3$. We recall that the CS matrix model on $\IS^3$, first considered in \cite{mm}, 
is defined by the partition function 
\be
Z_{\IS^3}(N,g_s)={1\over N!} \int \prod_{i=1}^{N}{ \rd \mu_i  \over 2\pi}  \prod_{i<j} \left( 2 \sinh  \left( {\mu_i -\mu_j \over 2}\right) \right)^2  \re^{-{1\over 2g_s} \sum_i \mu_i^2 }.
\ee
This is a one-cut matrix model \cite{leshouches}. It can be obtained from the lens space matrix model when one of the two cuts collapses to zero size. In the $Z$ plane the endpoints of the cut are given by $a$ and $a^{-1}$, where
\be
a=2\re^t -1-2 \re^{t/2} {\sqrt{\re^t-1}}.
\ee
Let us consider the following 
operator in this model:
\be
\CW(\nu_j)= 2 \sum_{i,j} \log \left( 2  \cosh  \left( {\mu_i -\nu_j \over 2}\right) \right). 
\ee
The lens space partition function (\ref{intdef}) can be calculated in two steps. In the first step, we compute 
\be
Z_1(\nu_j )= \left\langle  \re^{\CW(\nu_j)} \right\rangle_{N_1} 
\ee
where the subindex $N_1$ indicates that this is an unnormalized
VEV in the $\IS^3$ CS matrix model with gauge group $U(N_1)$. In a second step, we calculate 
\be
Z_{L(2,1)}=\langle Z_1(\nu_j) \rangle_{N_2}
\ee
in the CS matrix model with gauge group $U(N_2)$. To obtain the conifold expansion, we calculate $Z_1(\nu_j)$ and we expand it in $g_s$ and around $\nu_j=0$. Each 
term in this expansion can be computed exactly as a function of the K\"ahler parameter $t_1$, 
since the CS matrix model can be solved exactly in the $1/N$ expansion. The resulting 
double series in $g_s$ and $\nu_j$ is then regarded as an operator in the 
CS matrix model with group $U(N_2)$, which we expand around the Gaussian point as in \cite{mm,akmv}, {\em i.e.}, we expand the $\sinh$ measure around $\nu_j=0$. The 
partition function $Z_{L(2,1)}$ is then computed as a VEV in the Gaussian matrix model. This procedure gives a method to compute the expansion 
(\ref{fgconex}) directly in the matrix model.

To illustrate this procedure, let us calculate $F_0(t_1, t_2)$ at first order in $t_2$. In this computation we will denote 
\be
U_1=\text{diag} (\re^{\mu_i}), \qquad U_2=\text{diag} (\re^{\nu_j}).
\ee
The expansion around $\nu_j=0$ of the operator $\CW(\nu_j)$ reads
\be
\CW(\nu_j) =2 N_2 \sum_{i=1}^{N_1} \log \left[ 2 \cosh\left( {\mu_i \over 2}\right)\right]-
\sum_{j=1}^{N_2} \nu_j \sum_{i=1}^{N_1} \tanh\left( {\mu_i \over 2}\right) +\CO(\nu_j^2).
\ee
The average of the second term in the $U(N_2)$ matrix model vanishes (since it is odd in $\nu_j$), while higher order terms are at least of order $t_2^2$. The first term can be 
written as
\be
\label{ovope}
 2 \sum_{i=1}^{N_1} \log \left[ 2 \cosh\left( {\mu_i \over 2}\right)\right]
 = 2 \, \tr \, \log(1+U_1)- \sum_{i=1}^{N_1} \mu_i .
\ee
Therefore, in the planar limit and neglecting terms which contribute at order $t_2^2$, we have
\be
\label{expoZ}
\log Z_1(\nu_j) \approx {2 t_2\over g_s} \left  \langle \tr \, \log(1+U_1) \right\rangle_{N_1} 
\ee
since the second term in (\ref{ovope}) is odd in $\mu_i$ and its VEV vanishes. We then find,  
\be
\label{fccorr}
F_0(t_1, t_2) = F_0^{\IS^3}(t_1) + 2 t_2  g_s \left  \langle  \tr \, \log(1+U_1) \right\rangle +\CO(t_2^2).
\ee
The VEV in (\ref{fccorr}), which is now normalized, can be computed in terms of the resolvent of the CS matrix model, and similar computations appear in \cite{abms,mpp} in the context 
of large $N$ instanton corrections. In fact, it follows from (\ref{defgY}) and (\ref{fullintegral}) that the VEV in (\ref{fccorr}) is given by $-g (-1)$, where $g(Y)$ is computed in (\ref{gS3ex}). The final result for the linear correction in $t_2$ is 
\be
{\pi^2 \over 3} +{t_1^2\over 2} + {\rm Li}_2(\re^{-t_1}) - 2 {\rm Li}_2(\re^{-t_1/2}) +2 {\rm Li}_2(-\re^{-t_1/2}). 
\ee
Using dilogarithm identities, this agrees with $\frac{\lambda_2}{t_2}F_0^{(1)}(\lambda_1)$ in (\ref{conlist}). 
It is interesting to point out that, in the context of CS theory on the lens space $L(2,1)$, this function is essentially the action of the large $N$ instanton corresponding to the flat connection 
\be
U(N) \rightarrow U(N_1) \times U(N_2), \qquad N_2 \ll N_1,
\ee
as shown in \cite{mpp}. In the matrix model, this action is obtained by tunneling $N_2$ eigenvalues from the first cut to the second cut. 

We can also calculate the conifold expansion for the VEV of $1/6$ and $1/2$ BPS Wilson loops directly in the matrix model. We want to compute 
\be
\langle W_{\tableau{1}}^{1/6}\rangle =g_s\langle \tr \, U_1 \rangle_{L(2,1)}. 
\ee
We will again perform this computation in the planar approximation and at linear order in $t_2$. At this order we can compute instead the normalized average of the operator
\be
{\left\langle \tr  \,  U_1 \, \re^{\CW(\nu_j)}\right\rangle_{N_1} \over \left\langle \re^{\CW(\nu_j)}\right \rangle_{N_1}} =\langle \tr\, U_1\rangle + \left\langle \tr \, U_1 \, \CW(\nu_j)\right\rangle^{(c)}+ \cdots
\ee
in a Gaussian matrix model for the $\nu_j$. In the last line, all VEVs are normalized VEVs in the $\IS^3$ CS matrix model. By completing the square of the Gaussian weight 
we derive
\be
\left\langle \tr \, U_1 \left( \sum_{i=1}^{N_1} \mu_i \right)\right\rangle={\partial \over \partial j}\left\langle  \tr \, U_1 \, \re^{j \sum_{i=1}^{N_1} \mu_i} \right\rangle \biggl|_{j=0}=g_s\left\langle \tr \, U_1 \right\rangle .
\ee
We then find, at this order, 
\be
\label{tocom}
\langle W_{\tableau{1}}^{1/6}\rangle_{g=0}=g_s \langle \tr \, U_1\rangle +t_2  \left(  2 \langle \tr \, U_1 \, \tr \log (1+ U_1) \rangle^{(c)}- g_s \langle \tr \, U_1\rangle \right) + \CO(t_2^2). 
\ee
The connected correlator  
\be
\langle \tr \, U_1 \, \tr \log (1+ U_1) \rangle^{(c)}=-\sum_{\ell=1}^{\infty} {(-1)^{\ell} \over \ell} \langle \tr \, U_1 \, \tr \,U_1^{\ell} \rangle^{(c)}
\ee
can be computed by considering the (partially) integrated two-point function (see for example \cite{ambjorn})
\be
\int \rd p \, W_0(p,q)=-\sum_{n,m} {1\over n p^n q^{m+1}} \langle \tr\, U_1^{n} \, \tr\, U_1^{m} \rangle^{(c)}
\ee
and extracting the coefficient of $q^{-2}$. We have, 
\be
\int \rd p \, W_0(p,q)={1\over 2(p-q)} \left( 1- {\sqrt{ {(p-a)(p-a^{-1} ) \over (q-a)(q-a^{-1})}}}\right)+  {1\over 2\sqrt{(q-a)(q-a^{-1})}}, 
\ee
which includes the appropriate integration constant. We find, after changing $p \rightarrow -p$, 
\be
-\sum_{\ell=1}^{\infty} {(-1)^{\ell} \over \ell p^{\ell}} \langle \tr\, U_1 \, \tr\, U_1^{\ell} \rangle^{(c)} ={1\over 4}\left( a+a^{-1} + 2p -2 {\sqrt{(p+a)(p+a^{-1})}}\right).
\label{lsum}
\ee
When $p=1$ this gives
\be
 -\sum_{\ell=1}^{\infty} {(-1)^{\ell} \over \ell } \langle \tr\, U_1 \, \tr\, U_1^{\ell} \rangle^{(c)} =\re^{t_1} -\re^{t_1/2}.
\ee
Notice that this is an infinite sum of correlators in the CS matrix model. 
Since
\be
\langle \tr\, U_1\rangle ={\re^{t_1} -1\over g_s}, 
\ee
we finally obtain, 
\be
\label{1/6-t2}
\ba
 \langle W_{\tableau{1}}^{1/6}\rangle_{g=0}&=\re^{t_1} -1  +t_2  \left( \re^{t_1/2}-1\right)^2 +\CO(t_2^2) \\
 &=
\re^{t_1/2}\left(2\sinh\frac{t_1}{2}
+t_2\left(-2+2\cosh \frac{t_1}{2}\right) +\CO(t_2^2)\right).
\ea
\ee
Since this is a Wilson loop only in the first group, the framing prefactor depends only 
on the first 't~Hooft coupling.

The $1/2$ BPS Wilson loop is obtained by subtracting 
\be
 \langle \tr \, U_2 \rangle_{L(2,1)} =N_2 +\CO(t_2^2)={t_2\over g_s} + \CO(t_2^2). 
\ee
We find, 
\be
\re^{-(t_1 +t_2)/ 2}  \langle W_{\tableau{1}}^{1/2}\rangle_{g=0}= 2 \sinh\left( {t_1\over 2}\right) +t_2 \left( -2+ \cosh\left( {t_1\over 2}\right) \right) +\CO(t_2^2).
\ee
This is the result (\ref{12coni}) obtained from the conifold expansion after using the dictionary (\ref{relthoofts}).

\subsection{On the near Chern--Simons expansion of ABJM theory}
\label{sec:near-CS}

In the matrix model the conifold locus corresponds to vanishing of one of the two cuts, 
where the lens space matrix model can be written as a perturbation around the matrix 
model for Chern--Simons on $\IS^3$. Here we want to explore this limit in the original 
3-dimensional theory.

In the strict limit we have the theory with $N_2=0$ and $N_1\gg1$ and arbitrary 
$N_1/k$. In this limit all the fields charged under the second gauge group, {\em i.e.}, its 
gluons and all the bi-fundamental fermions and scalars are removed. Consequently, 
ABJM theory simplifies dramatically and reduces to topological $U(N)$ CS. The only 
observables in the theory in this strict limit are Wilson loops, and they are given by 
the standard CS answer, which is exact in $\lambda_1$ (and $1/N_1$).

One can try to perform a systematic expansion around this point in a perturbative 
expansion in $\lambda_2$. One keeps $\lambda_2\ll\lambda_1$, but if desired, 
can still assume the planar approximation, ignoring also the $1/N_2$ corrections.

It is convenient to draw the Feynman graphs in double-line double-color notation, 
one color for each group. At the first non-trivial order in $\lambda_2$, only graphs 
with a single index loop of $U(N_2)$ are included. An arbitrary number of gluons 
of $U(N_1)$ are allowed. Let us propose the following calculation procedure: 
First ignore all $U(N_1)$ gluons and enumerate all remaining graphs. They 
are a very restricted subset, which can be identified very easily.

In order to dress them up with the $U(N_1)$ gluons we write the proparators for the 
bi-fundamental fields as a path integral over all trajectories in space. 
As charged object, these paths will effectively be Wilson loop in $U(N_1)$, which 
can be calculated exactly in CS theory. Since this theory is topological, the result 
of adding all the gluons, is independent of the path of the bi-fundamental fields. 
One can then do the usual path integral for these fields and find the regular scalar 
and fermion propagators.

The statement in the previous paragraph fails in a subtle way. The correlation 
function of Wilson loop operators in CS theory does not depend on their geometry only 
as long as their topology --- the knotting and linking numbers --- are kept fixed. 
Therefore one has to modify the above statement, and sum over all possible 
topologies of the paths of the bi-fundamental fields accompanied by the relevant 
knotted/linked Wilson loops. Unfortunately, we do not have an {\em a-priori} method 
of determining the weight that should be assigned to the different topologies.%
\footnote{Wilson loops arise out of dressing 
propagators of matter fields also in \cite{aekms}. In that case the path is fixed to 
a collection of light-like segments, due to the singularity in the Minkowski-space 
propagator.}

As an illustration, let us consider the $1/6$ BPS Wilson loop (whose Feynman 
rules are simpler than the $1/2$ BPS one) and examine its perturbative 
expansion about the conifold locus. 
The Wilson loop is given in our normalization by \cite{dp,cw,rey}
\be
W^{1/6}_{\tableau{1}}=g_s\,\tr\,\CP\exp \int\left(\ri A_\mu \dot x^\mu
+\frac{2\pi}{k}|\dot x|M^I_J C_I \bar C^J\right)ds.
\label{expl-wl}
\ee
$x^\mu$ parameterizes a circle in $\IR^3$ (or $\IS^3$), $A_\mu$ are the 
$U(N_1)$ gluons, $C_I$ and $\bar C^I$ are the bi-fundamental scalars and 
$M^I_J=\text{diag}(1,1,-1,-1)$ is a matrix in flavor space, which is required to make 
this object BPS.

At order $\CO(\lambda_2^0)$, this is simply a Wilson loop of CS, whose planar 
expectation value (ignoring framing) is
\be
\langle W^{1/6}_{\tableau{1}}\rangle_{g=0}=2\ri\sin\pi\lambda_1+\CO(\lambda_2).
\ee

\FIGURE{
\includegraphics[width=15.5cm]{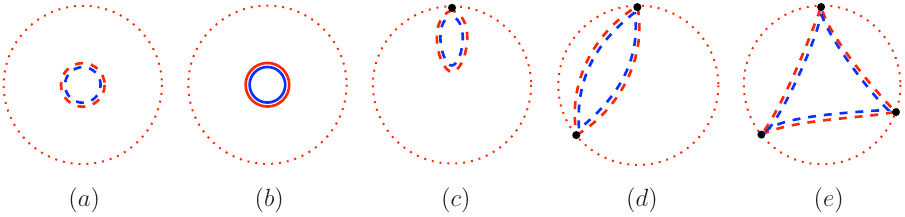} 
\caption{Several Feynman graphs which may contribute at order $\lambda_2$ to the 
$1/6$ BPS Wilson loop with gluons stripped. The big circle is the Wilson loop, 
dashed lines are bosons and the solid line a fermion, all presented in double-line, 
double-color notation.}
\label{wl-pert}
}

After stripping away the gluon lines, there are still an infinite number of graphs involving 
bi-fundamental fields. Examples are shown in \figref{wl-pert}. In the example drawn, 
there is a single scalar or fermion loop. The scalar loop can ``touch'' the Wilson loop at an 
arbitrary number of points, due to the scalar bilinear term in (\ref{expl-wl}). There are 
extra graphs which are not drawn, with fermionic tadpoles on the scalar lines, or 
vice--versa.

By explicit calculations \cite{dp,cw,rey}, all the connected graphs illustrated 
(\figref{wl-pert}$c$, \ref{wl-pert}$d$, \ref{wl-pert}$e$) vanish in 
dimensional regularization. The same can be argued for higher order graphs of this form. 
Likewise, one would not expect tadpoles to contribute. We are left therefore with the 
first two disconnected graphs, which become connected once gluon lines are added. 
Indeed, the only non-vanishing graph that was thus far calculated is the one-loop correction 
to the gluon propagator (\figref{wl-pert}$a$, \ref{wl-pert}$b$ with two extra gluons), 
which accounts for the $\CO(\lambda_2\lambda_1^2)$ term 
(which with our normalization is 2-loops) 
in the explicit answer (\ref{16weak}).%
\footnote{This graph has a divergence that can be removed by including the double 
scalar exchange graph (\figref{wl-pert}$c$). In dimensional regularization the 
finite part comes only from the gluon graph.}

We can compare this to the explicit calculation in the matrix model above. The essential 
part of the expression for the Wilson loop at order $\CO(\lambda_2)$ (\ref{tocom}) 
is the connected correlator of two Wilson loops. One of them is the original Wilson loop 
and the other came from expanding the $\cosh$ term in the matrix model (\ref{ovope}), 
which arises from integrating out the bi-fundamental matter. So this agrees with the 
identification of the contribution as coming from the bubble graphs. Moreover, what 
we see in the matrix model is that one should sum over multi-winding of this second 
Wilson loop, with a weight $1/l$. This corresponds in the physical theory to summing 
over all possible topologies for the scalar and fermion bubble. As mentioned, we 
do not know how to derive this factor of $1/l$ from perturbation theory, but it is given 
to us by the explicit matrix model calculation.

It was noted in \cite{mss} that in this limit of ABJM theory 
the spectrum of local operators also simplifies and the spin-chain hamiltonian becomes 
short-range. A compelling conjecture for the mysterious function $h(\lambda)$ in that 
limit was also presented there.
It would be interesting to explore this limit further and learn how to do this 
sum over topologies for other observables.

\sectiono{Modular properties and the genus expansion}
\label{sec:modular}

In this section we provide an efficient, recursive method to compute the 
$1/N$ corrections to the free energy in the case $N_1=N_2=N$. This is based on the modular properties of the solution and the technique of 
direct integration of the holomorphic anomaly equations. The method determines {\it a priori} the full $1/N$ expansion. In practice it is quite 
efficient and it makes possible to calculate the $F_g$ corrections for high genera. This is then used to estimate non-perturbative effects in the large $N$ expansion.

As noted in \cite{abk}, we can use the relation between the local $\IF_0$ theory and Seiberg--Witten theory to write all the quantities in the model in terms of modular forms. 
This representation becomes particularly useful when we restrict ourselves to a one-parameter model, as it was shown in a different context in \cite{kmr}. When $N_1=N_2$, $\beta=1$ and the modulus $u$ becomes simply
\be
 u=1+{\kappa^2\over 8}. 
\ee
 In Seiberg--Witten theory, $u$ is related to the modular parameter $\tau$ of the Seiberg--Witten curve by
\be
\label{eorbu}
u=\frac{\vartheta^4_4-\vartheta^4_2}{\vartheta^4_3}(\tau)=1-32 q^{1/2} +256 q+\cdots
\ee
where $q=\re^{2 \pi \ri \tau}$. This formula can be inverted to 
\be
\label{tauex}
\tau=\ri  {K'\left({\ri \kappa \over 4}\right)\over K \left({\ri \kappa \over 4}\right)},
\ee
therefore we see that the modular parameter $\tau$ is related to the specific heat of the theory through (\ref{seconder}). 
Let us now introduce the quantity 
\be
\xi={2\over \vartheta_2^2(\tau) \vartheta_4^4(\tau)}.
\ee
This is proportional to the third derivative of the genus zero free energy, therefore to the Yukawa coupling $C_{\lambda \lambda \lambda}$. More precisely, we have
\be
\partial_\lambda^3 F_0(\lambda)= -8 \pi^3 \ri \, \xi. 
\ee
Therefore, the planar content of the theory can be elegantly encoded in terms of modular forms on the Seiberg--Witten curve.  

One powerful application of the modular properties of the ABJM theory is the determination of the higher genus corrections to the 
free energy, $F_g(\lambda)$. 
These can be obtained in principle from the matrix model (\ref{kapmm}), or equivalently from the formalism of \cite{eo} (appropriately modified as in \cite{mmopen,bkmp}). 
However, as emphasized in for example \cite{hk,hktwo,kmr}, this formalism is not very convenient to do calculations at higher genus. One should rather use the fact that the $F_g$ are 
quasi-modular forms that can be promoted to non-holomorphic modular forms. The resulting non-holomorphic objects satisfy the holomorphic anomaly equations of \cite{bcov}, 
as shown in \cite{hk,emo}, and these can be in turn solved with the technique of direct integration developed in \cite{hk,gkmw,hkr,kmr} for local CY manifolds and matrix models. 

The basic strategy of direct integration is the following. First, we assume an ansatz for $F_g$ of the form
\begin{equation}
F_g (\tau)=\xi^{2g-2} f_g (\tau) 
\ee
where 
\be
f_g(\tau)=\sum_{k=0}^{3g-3}
E_2^{k}(\tau) c^{(g)}_k(\tau)\ ,\qquad g\ge 2,
\label{eq:generallocalform}
\end{equation}
is an almost modular form of weight $6g-6$, with respect to a monodromy group 
$\Gamma \subset SL(2, \IZ)$. $F_g(\tau)$ can be promoted to a 
non-holomorphic modular form $F_g(\tau, \bar \tau)$ by changing
\be
E_2(\tau) \rightarrow \widehat E_2 (\tau,\bar \tau) =E_2(\tau)-{3\over \pi\, \text{Im}(\tau)}. 
\ee
The resulting $F_g(\tau, \bar \tau)$ satisfies the holomorphic anomaly equations of \cite{bcov}, which govern their anti--holomorphic dependence. Since 
this dependence is contained in $\widehat E_2(\tau, \bar \tau)$, these equations govern the $E_2$ content of $F_g$. 
This means that the coefficients $c^{(g)}_k(\tau)$, which are modular forms of weight $6g-6-2k$, can be obtained recursively for $k>0$ if one knows the lower 
$F_g$. In order to write down the recursive equation, it is useful to introduce a covariant derivative $\rd_\xi$ taking a form of weight $k$ into a form of weight $k+2$:
\begin{equation}
 \rd_\xi=\d_\tau+\frac{k}{3}\,\frac{\d_\tau \xi}{\xi}
\end{equation} 
Then, the holomorphic anomaly equations lead to
\begin{equation}
 \frac{\rd f_g}{\rd E_2}=-\frac{1}{3}\,\left\{\rd^2_\xi f_{g-1}+\frac{1}{3}\,\frac{\d_\tau \xi}{\xi}\,\rd_\xi f_{g-1}
 +\sum_{r=1}^{g-1}\rd_\xi f_r\,\rd_\xi f_{g-r}\right\}, \qquad g\ge 2.
\end{equation} 
If $F_{g'}$ are known, with $g'<g$, the above equation determines all the coefficients $c^{(g)}_k(\tau)$ in $f_g$, with the exception of $c^{(g)}_0(\tau)$, which 
plays the r\^ole of an integration constant. This coefficient is a holomorphic form of weight $6g-6$ and it is called the {\it holomorphic ambiguity}. 

In order to fix the holomorphic ambiguity we need two pieces of information. The first one concerns its functional dependence. 
Since $c^{(g)}_0(\tau)$ is a modular form w.r.t. some monodromy subgroup, 
it belongs to a finitely generated ring. This means that it is determined by a finite number of coefficients, which typically grows with $g$. 
The second piece of information comes from boundary conditions at singular points in moduli space. A very powerful boundary condition for matrix models and 
local Calabi--Yau manifolds is the so-called {\it gap condition}, discovered in \cite{hk} and further used in \cite{hkr,kmr} to fix the holomorphic 
ambiguity. According to the gap condition, near certain points $p_i$ in moduli space, parametrized by a flat coordinate $t_i$, the genus $g$ free energy behaves as 
\be
F^{(i)}_g ={a_g \over t_i^{2g-2}} + \CO(1). 
\ee
The superscript $(i)$ means that the genus $g$ free energy has to be transformed to the duality frame which is appropriate for the $i$-th singularity, as it is well-known in 
special geometry. The ``gap" refers to the absence of singular terms $t^{-k}$ with $0<k< 2g-2$ in the local expansion near $t_i=0$. The vanishing of these terms provides boundary conditions for $c^{(g)}_0(\tau)$, and in some cases it fixes them completely. 

In our case, the relevant ring is that of $\Gamma_2$ modular forms which is 
generated by the theta functions
\be
b=\vartheta_2^4(\tau)\;,\;\;c=\vartheta_3^4(\tau)\;,\;\;d=\vartheta_4^4(\tau). 
\end{equation} 
Since $c=b+d$, only two of them are independent, and we will choose $b$ and $d$. Using standard formulae in the theory of modular forms, one finds
\be
\frac{\d_\tau \xi}{\xi}=\frac{b-E_2}{4},
\ee
as well as
\be
 \rd_\xi b=\frac{b^2+bd}{3}, \qquad 
\rd_\xi (bd)=\frac{(bd)b}{6}, \qquad \rd_\xi E_2=\frac{1}{12}\left(-E_2^2+2bE_2-E_4\right). 
\ee
The modular expression for the genus one free energy is known \cite{abk} and reads
\be
F_1=-\log \eta(\tau), 
\ee
therefore we have 
\be
 \rd_\xi f_1 =-\frac{E_2}{24}.
\ee
These are all the ingredients needed for the recursion. The holomorphic ambiguity can be written as
\be
\label{holex}
 c^{(g)}_0(\tau)=\sum_{j=0}^{3g-3} \alpha_j^{(g)} b^j d^{3g-3-j}
\ee
and it involves $3g-2$ unknowns. Let us see how we can fix these by looking at the behavior near the three singular points of moduli space.

At the orbifold point, the $F_g$ are the genus $g$ amplitudes of the super-matrix model (\ref{kapmm}) with $N_1=N_2$. Their leading behavior near $\lambda=0$ is governed by two copies of the Gaussian matrix model, therefore they behave as
\be
\label{oex}
F^{( \text{o})}_g(\lambda)= {B_{2g} \over g (2g-2)} (2 \pi \ri \lambda)^{2-2g} + \CO(\lambda^2). 
\ee
This gives $g$ conditions, since the ansatz (\ref{holex}) for the holomorphic ambiguity only involves 
even powers of $\lambda$.  

The symmetric conifold point $z_1=z_2=1/16$ is related to the orbifold point through an $S$-duality transformation. The appropriate global coordinates near this point are 
given in (\ref{conglobal}). In the ABJM slice one has
\be
y_1=0,\qquad y_2=y=1-\frac{\zeta^2}{16}.
\ee
The following period is a good local, flat coordinate near the symmetric conifold point:
\be
 t=\sum_{n=0}^\infty\frac{a_n}{(n+1)\,2^{4n}}y^{n+1}, 
\ee
where
\be
a_n={1\over {2n \choose n}} \sum_{k=0}^n{2k \choose k}{4k \choose 2k}{2n-2k \choose n-k}{4n-4k \choose 2n-2k}.
\ee
It was noticed in \cite{hkr} that the genus $g$ amplitude at the conifold point behaves like
\be
\label{cex}
F^{( \text{c})}_g(t)= {B_{2g} \over 2g (2g-2)} \left( {t \over 2 \ri}\right)^{2-2g} + \CO(1). 
\ee
This fixes $2g-2$ conditions. Together with the $g$ conditions coming from the orbifold point, this completely fixes the $3g-2$ unknowns in the holomorphic ambiguity. 

The result can be verified by looking at the radius point, which is related to the orbifold point by an $STS$ transformation. 
The genus $g$ free energy at this point is the generating function of Gromov--Witten invariants 
of the local $\IF_0$ geometry in the slice $T_1=T_2=T$. More precisely, one has
\be
\label{lrex}
F^{( \text{GW})}_g(Q)  =(-4)^{g-1} \left\{ {(-1)^g |B_{2g} B_{2g-2}| \over g (2g-2) (2g-2)!} + \sum_{d\ge1} N_{d,g} Q^d \right\}, \qquad Q=\re^{-T}
\ee
where 
\be
N_{d,g}=\sum_{d_1+d_2=d} N_{d_1, d_2,g}
\ee
is a sum of Gromov--Witten invariants at genus $g$, $N_{d_1, d_2, g}$, of local $\IF_0$ (the degrees $d_1, d_2$ correspond to the two 
K\"ahler classes of this geometry). The constant term in (\ref{lrex}) is the well-known constant map contribution to the higher genus free energy \cite{bcov} for a manifold 
with ``effective" Euler characteristic $\chi=4$. It can be checked 
that the higher genus free energies obtained from the integration of the holomorphic anomaly equation with the above boundary conditions reproduce the well-known 
large radius free energies (\ref{lrex}).  

Let us see how this works in some detail when $g=2$. The integration of the holomorphic anomaly equation gives, 
\be
 f_2=\frac{1}{3}\cdot\frac{1}{24^2}\left(-\frac{5}{3}\,E_2^3+3bE_2^2-2E_4E_2\right) + c^{(2)}_0 (\tau), 
\ee
where $c^{(2)}_0 (\tau)$ is of the form (\ref{holex}). The expansion around the orbifold and conifold points read, respectively,  
\be
\ba
F^{( \text{o})}_2 (\lambda)&=\frac{1}{432 (2\pi \ri \lambda)^2}\left(-\frac{11}{3}+1728 \alpha^{(2)} _0 \right)+4 \left(\frac{2}{3} \left(\alpha^{(2)}_0-\frac{11}{5184}\right)-\frac{3\alpha^{(2)}_0}{2}+\alpha^{(2)}_1+\frac{1}{576}\right)\\
&+\CO(\lambda^2),\\
F^{( \text{c})}_2 (t)  &=-\frac{5+1296 \alpha^{(2)} _3}{1296 t^2}+\frac{-1-864(12 \alpha^{(2)} _2+15 \alpha^{(2)} _3)}{10368 t}+\CO(1),\\
\ea
\ee
Imposing the conditions (\ref{oex}), (\ref{cex}) and (\ref{lrex}) we fix 
\be
\alpha^{(2)}_0={1\over 25 920}, \quad \alpha^{(2)}_1=-{1 \over 3456}, \quad \alpha^{(2)}_2={1\over 3456}, \quad \alpha^{(2)}_3={1\over 3240}. 
\ee
We finally obtain
\be
F_2^{(\text{o})}=\frac{1}{432 b d^2}\left(-\frac{5}{3}\,E_2^3+3bE_2^2-2E_4E_2\right)+\frac{16 b^3+15 d b^2-15 d^2 b+2 d^3}{12960 b d^2}, 
\ee
which gives at large radius the expansion 
Since $\tau$ depends on $\lambda$ through (\ref{tauex}) and (\ref{lamkap}), this gives the exact expression for the genus two free energy on $\IS^3$ in the ABJM 
model, for any value of the 't~Hooft coupling. 

Notice that the modular ring appearing here and parametrizing the holomorphic ambiguity is different from the one appearing in Seiberg--Witten theory \cite{hk,gkmw} or in the 
cubic matrix model \cite{kmr}. This is 
due to the fact that, although the curves are the same, the meromorphic forms defining the theory are different. 

Using this method, we have computed the free energies up to high genus. The strong coupling behavior of $ F_g^{(\rm o)}$ is of the form 
\be
 F_g^{(\rm o)}(\lambda)-c_g \sim -\lambda^{{3\over 2}-g}, \qquad \lambda \to \infty, \quad g\ge0,  
\ee
where 
\be
c_g=-  { 4^{g-1} |B_{2g} B_{2g-2}| \over g (2g-2) (2g-2)!} 
\ee
is the constant map contribution appearing in (\ref{lrex}). 
We have also used these results in order to investigate the large order behavior of the $1/N$ expansion. We have found that 
\be
\label{signF}
\widetilde F_g (\lambda) =(-1)^{g-1} \left( F^{(\rm o)}_g- {B_{2g} \over g (2g-2)} (2 \pi \ri \lambda)^{2-2g} - c_g \right) 
\ee
behaves at large $g$ as 
\be
\label{loeffect}
\widetilde F_g(\lambda) \sim (2g)! \left|A(\lambda)\right|^{-2g} \cos\left( 2g\theta(\lambda) +\delta(\lambda) \right).
\ee
In this equation, the angle $\theta(\lambda)$ satisfies $\theta(0)=\pi/2$ and $\theta(\lambda)\not=0$ for all $\lambda>0$, while $\delta(\lambda)$ is a function of $\lambda$ 
(see for example section 2 of \cite{msw} for more details on the large order behavior of the genus expansion). 
The sign $(-1)^{g-1}$ is included in (\ref{signF}) since in the physical ABJM theory the 
coupling $g_s$ is imaginary. The large order behavior (\ref{loeffect}) indicates that 
the singularities of the Borel transform of $F_g(\lambda)$ which are closest to the origin are located at $\pm A(\lambda)$, where
\be
A(\lambda)= \left|A(\lambda)\right| \re^{\ri \theta(\lambda)}.
\ee
Since $\theta(\lambda)$ does not vanish, none of them lies on the positive real axis. This strongly suggests 
that the $1/N$ expansion of the free energy is Borel summable for any $\lambda>0$.  

The large order behavior of the genus expansion (\ref{loeffect}) is similar to the one found for 
Chern--Simons theory on $\IS^3$ in \cite{ps}, and it should be governed by a large $N$ instanton with action $A(\lambda)$. It would be very interesting to identify this 
instanton and compute $A(\lambda)$ analytically, both in the gauge theory and in the string theory dual. 
The factorial growth, found here by explicit calculation in the matrix model, agrees with the expected 
behavior for the genus expansion in string theory \cite{shenker}.

\sectiono{More exact results on Wilson loops}
\label{sec:wilson}

In this section 
we elaborate on the results of \cite{dt,mp} and we obtain more exact results on Wilson loops. 

\subsection{$1/N$ corrections}
The higher genus corrections to the VEV of $1/2$ and $1/6$ BPS Wilson loops  can be computed in terms of the higher genus corrections to the resolvent of the 
matrix model. The resolvent has a genus expansion of the form 
\be
\omega(z) =\sum_{g=0}^{\infty} g_s^{2g}\omega_g(z). 
\ee
In the same way, the density of eigenvalues has a large $N$ expansion of the form 
\be
\rho(z)=\sum_{g=0}^{\infty} g_s^{2g}\rho_g(z)\,,
\qquad
\rho(z)=\rho^{(1)}(z)+\rho^{(2)}(z).
\ee
The $\rho^{(i)}_g(z)$ (with $i=1,2$) have their support on the intervals $\CC_{i}$, 
and they can be obtained by the discontinuity of $\omega_g$ at the cuts as in 
(\ref{gdensities}).

The genus expansion of the expectation value of the $1/6$ BPS and $1/2$ BPS 
Wilson loops follows the expressions in (\ref{1/6-integral}) and (\ref{1/2-integral}) with 
the appropriate term in the expansion of $\rho^{(i)}(Z)$ and $\omega(Z)$.

The first step is therefore to compute $\omega_g(p)$. This calculation 
can be done with the recursive techniques developed in the matrix model literature 
starting with \cite{ambjorn} and culminating with \cite{eo}. We will perform an explicit 
computation for $g=1$. Calculations for $g\ge2$ are in principle doable, but they 
become complicated. 

A convenient formula for $\omega_1(p)$ for an algebraic resolvent was found in \cite{akemann}. To write this formula, we write the discontinuity of the resolvent (also called {\it spectral curve} in the 
matrix model literature) as 
\be
\label{resmom}
 y(p)=M(p)\sqrt{\sigma(p)}, \qquad \sigma(p)=(p-x_1)(p-x_2)(p-x_3)(p-x_4). 
\ee
$M(p)$ is sometimes called the moment function. Then, one has
\be
\label{akeone}
 \omega_1(p)=\frac{4}{\sqrt{\sigma(p)}}\sum_{i=1}^4\left(\frac{A_i}{(p-x_i)^2}+\frac{B_i}{p-x_i}+C_i\right), 
\ee
where
\be
\label{abc}
\ba
 A_i&=\frac{1}{16}\,\frac{1}{M(x_i)}, \\
  B_i&=-\frac{1}{16}\,\frac{M'(x_i)}{M^2(x_i)}+\frac{1}{8\,M(x_i)}\bigg(2\alpha_i-\sum_{j\neq i}\frac{1}{x_i-x_j}\bigg),\\
   C_i&=-\frac{1}{48}\,\frac{1}{M(x_i)}\sum_{j\neq i}\frac{\alpha_j-\alpha_i}{x_j-x_i}-
 \frac{1}{16}\,\frac{M'(x_i)}{M^2(x_i)}\,\alpha_i+\frac{\alpha_i}{8\,M(x_i)}
 \bigg(2\alpha_i-\sum_{j\neq i}\frac{1}{x_i-x_j}\bigg),
 \ea
\end{equation}
and the $\alpha_i$ are given by 
\be
\ba
\alpha_1&={1\over(x_1-x_2)}\left[1-{(x_4-x_2)\over(x_4-x_1)}{E(k)\over K(k)}
\right],\cr
\alpha_2&={1\over(x_2-x_1)}\left[1-{(x_3-x_1)\over(x_3-x_2)}{E(k)\over K(k)}
\right],\cr
\alpha_3&={1\over(x_3-x_4)}\left[1-{(x_4-x_2)\over(x_3-x_2)}{E(k)\over K(k)}
\right],\cr
\alpha_4&={1\over(x_4-x_3)}\left[1-{(x_3-x_1)\over(x_4-x_1)}{E(k)\over K(k)}
\right], 
\ea
\ee
where the modulus of the elliptic functions is 
\be
k^2={(x_1-x_2)(x_3-x_4)\over(x_1-x_3)(x_2-x_4)}.
\ee
These expressions differ from the ones in \cite{akemann} in a permutation of the roots, as explained in \cite{kmt}. The overall factor of $4$ in (\ref{akeone}) is 
due to the fact that our resolvent has a different normalization than the one in \cite{akemann}. 

Although the resolvent of the lens space matrix model (\ref{explicitRes}) is not algebraic, its discontinuity can be written in the form (\ref{resmom}) with 
\be
 \sigma(p)=f(p)^2-4\beta^2p^2, \qquad
 f(p)=p^2-\zeta p+1
\end{equation} 
and
\begin{equation}
 M(p)=\frac{2}{p\sqrt{\sigma(p)}}\tanh^{-1}\frac{\sqrt{\sigma(p)}}{f(p)}.
\end{equation}
This form of the spectral curve is typical of the mirrors of toric geometries \cite{mmopen,bkmp}. The branch points are
\begin{equation}
 x_1=-b, \quad x_2=-{1\over b}, \quad x_3={1\over a}, \quad x_4=a.
\end{equation} 
Using these expressions, it is possible to compute the integral 
\begin{equation}
\langle W_{\tableau{1}}^{1/6} \rangle_{g=1}=\frac{1}{4\pi \ri}\,\oint_{\CC_1} \omega_1(Z)Z \rd Z
\end{equation}
in closed form, in terms of elliptic functions $E, K$ and the elliptic integral of the third kind $\Pi(n,k)$, with 
\begin{equation}
n=\frac{(a^2-1)b}{(1+ab)a}\;\;.
\end{equation}
One finds the rather complicated expression
\be
\ba 
\langle W&_{\tableau{1}}^{1/6}\rangle_{g=1}=
\frac{1}{12 \pi\sqrt{a}\,{b}^{3/2} (1+ab)(a^2-1)^2(b^2-1)K}
\biggl[-3 (b-2a +a^2 b)\, (1+a b)^4E^2
+\Bigl[a (1+a^4)
\\&
-b+a^2 (4+4 a^2-a^4) b- 4a (1-3 a^2+a^4) b^2
-a^2(1+a^2)b^3 (2+b^2) +a(1-8 a^2+a^4) b^4\Bigr] K^2
\\ 
& +\left(b^3(1+6 a^2+a^4) +4 a (1+a^2)(b^2 -1)+b(3-14 a^2+3 a^4) \right)\,(1+a b)^2  E\,K 
\biggr]
\\
&+\frac{(a b-1) (a^2-b^2)}{12 \pi\,(ab)^{3/2}(1+ab)k^4K^2}
\Big[-6 E^2+4 (2-k^2) E\,K-(2-2k^2+k^4)\,K^2\Big]\Pi.
\ea
\ee
To check this formula, we expand it 
around the weakly coupled point $\lambda_1=\lambda_2=0$. 
After using the inverse mirror map given by (\ref{weakkappa}) we find
\begin{equation}
\label{gonew}
\ba
\langle W_{\tableau{1}}^{1/6} \rangle_{g=1}=&-\frac{\pi\ri}{12}\lambda _1
+\frac{\pi^2}{12} \lambda _1^2+\frac{\pi^2}{4} \lambda _1 \lambda _2
 +\frac{\pi^3\ri}{18}  \lambda _1^3+\frac{\pi^3\ri}{24}  \lambda _1^2 \lambda _2-\frac{\pi^3\ri}{4} \lambda _1 \lambda _2^2 \\ 
& -\frac{\pi ^4}{36} \lambda _1^4+\frac{\pi^4}{24}  \lambda _1^3 \lambda _2+\frac{5\pi^4}{24}\lambda _1^2 \lambda _2^2-\frac{\pi^4}{6}\lambda _1 \lambda _2^3+\mathcal{O}(\lambda^5).
\ea
\end{equation}
We can test this expansion with a perturbative calculation in the ABJM matrix model. At order $\CO(g_s^4)$ we have found, 
\be
\ba
& {\re^{-g_s N_1/2} \over 2\pi\ri\lambda_1} \langle W^{1/6} _{\tableau{1}} \rangle=1+ \left( {1\over 24} N_1^2 -{1\over 4} N_1 N_2 - {1\over 24} \right)   g^2_s  + \left( {1\over 16}  N_1  N_2^2  -{1\over 16} N_2 \right)   g^3_s \\
&\qquad+  \left( {3\over 5760} N_1^4 -{10 \over 1920} N_1^3  N_2   -{20 \over 1920} N_1  N_2^3 -{10 \over  5760} N_1^2 + {5\over 192} N_1 N_2+{1\over 32} N^2_2+ {7 \over 5760}  \right)   g^4_s +\cdots
\ea
\ee
It is straightforward to see that this agrees with (\ref{gonew}). 

The $1/N$ correction to the $1/2$ BPS Wilson loop is much easier to obtain, since it can be computed as a residue at infinity. We have that 
\begin{equation}
\omega_1(Z)=\frac{4}{Z^2}\sum_{i=1}^4C_i+\mathcal{O}(Z^{-3}),
\end{equation} 
where the $C_i$ are given in (\ref{abc}). We find, at weak coupling,
\be
\ba
 \langle W^{1/2} _{\tableau{1}} \rangle_{g=1}=&-\frac{\pi\ri}{12}(\lambda _1+ \lambda _2)
+\frac{\pi^2}{12}( \lambda _1^2- \lambda _2^2) 
+\frac{\pi^3\ri}{18}( \lambda _1^3+\lambda _2^3) -\frac{5\pi^3\ri}{24}\lambda _1 \lambda _2( \lambda _1 +\lambda _2)
\\
&-\frac{\pi ^4}{36} ( \lambda _1^4 -\lambda_2^4) +\frac{5\pi ^4}{24}  \lambda _1 \lambda _2( \lambda _1^2 - \lambda _2^2) +\mathcal{O}(\lambda^5).
\ea
\ee
At strong coupling we find (we consider for simplicity the ABJM slice)
\be
\langle W^{1/2} _{\tableau{1}} \rangle_{g=1}
=\frac{1}{24 \ri}\,\frac{3+2\log^2\kappa-4\log\kappa}{\log^2\kappa} \kappa +\mathcal{O}(1)\;
\ee
The leading exponent is exactly as at genus zero (\ref{12strong}), representing 
the same minimal surface with an extra degenerate handle attached. Its effect 
is to modify the one-loop determinant, which (with our normalization and 
ignoring instantons) can 
be written as
\be
\langle W^{1/2} _{\tableau{1}} \rangle_{g=1} 
=-\ri\left(\frac{1}{12}-\left(\frac{1}{6  \pi} + {\pi \over 288} \right){1\over  \sqrt{2\lambda}}+\CO\left({1\over \lambda} \right)\right)
\re^{\pi\sqrt{2 \lambda}},
\qquad\lambda\to\infty.
\ee

\subsection{Giant Wilson loops}
\label{sec:giant}

It has been argued in \cite{df,yamaguchi,gpone,gptwo} that a D-brane probe in 
$\text{AdS}_5\times\IS^5$ represents an insertion of a Wilson loop in the dual 
4d $\CN=4$ SYM with a large 
symmetric or antisymmetric representation (in the case of D3 branes and 
D5 branes, respectively). These ``giant Wilson loops" are characterized by a 
representation with $n$ boxes, and one considers the limit
\be
\label{giantl}
n, \, N \to \infty, \qquad {n \over N}~~\text{fixed}. 
\ee
In terms of the Gaussian matrix model of the Wilson loops in that theory, 
the giant Wilson loop in the symmetric representation is represented by an additional 
eigenvalue outside the cut and the antisymmetric representation by a ``hole'' in the original 
cut.

Let us review now the known D-brane solutions which could be relevant for ABJM theory. 
The usual $1/2$ 
BPS Wilson loop in the fundamental representation is described by a string with world-volume 
$\text{AdS}_2\subset\text{AdS}_4$. In M-theory it is an M2-brane wrapping also the orbifold cycle 
on $\IS^7/\IZ_k$. When considering $k/2$ coincident M2-branes (or $k$, when it is odd) 
the M2-brane solution develops an extra branch, where the circle becomes a linear combination 
of the orbifold direction and a contractible circle in AdS$_4$ \cite{lunin}. 
In type IIA these configurations are D2-branes with world-volume 
$\text{AdS}_2\times\IS^1\subset\text{AdS}_4$, where the radius of the $\IS^1$ is a free 
modulus. From the M-theory point of view these are continuous deformations of the system 
of $k/2$ coincident M2-branes describing a Wilson loop in a $k/2$ dimensional representation. 
In the field theory they are the vortex loop operators of \cite{vortex}, which have a description as 
semi-classical field configurations and carry the same charge as $k/2$ Wilson loops.

These solutions have further moduli associated to rotations away from the orbifold cycle 
inside $\IS^7/\IZ_k$. Such M2-brane configurations 
preserve 8 supercharges ($1/3$ BPS) \cite{dp,vortex}. 

There is also a known family of D6-brane solutions which were argued in \cite{dp} 
to represent the $1/6$ BPS Wilson loops in anti-symmetric representations. The action 
for this D-branes is (for $N_1=N_2$)
\be
S_\text{D6}=-\pi\sqrt{2\lambda}\,\frac{n(N-n)}{N}\,,
\label{16giant}
\ee
which matches that of $n$ strings for small $n$ and has the $n\to N-n$ symmetry of the 
antisymmetric representation.
In the matrix model these D6-branes should correspond to creating a ``hole'' in one of 
the two cuts, splitting it in two.

We turn now to the lens space matrix model and try to find the appropriate description for these 
objects, and in particular the $1/2$ BPS vortex loop operators. 
As pointed out in \cite{hku}, the calculation of Wilson loops in the matrix model in this 
limit can be done in a saddle-point approximation. We will now 
reformulate the arguments of \cite{hku} and adapt them to the lens space matrix model.

We will focus on the case of $1/2$ BPS Wilson loops, where we want to calculate 
\be
\label{weta}
W_n^{\eta}=\langle \tr_{\CR^\eta_n} U \rangle, \qquad \eta=\pm 1, 
\ee
where $U$ is the same matrix as in (\ref{U}) and $\CR_n^{\pm 1}=S_n$, $A_n$ are 
respectively the totally symmetric and the totally antisymmetric representations of 
$U(N_1+N_2)$ with 
$n$ boxes. It will turn out that the relevant limit in this theory is slightly different from 
(\ref{giantl}) and is given by fixing
\be
\label{nu}
\nu=\eta\,\frac{n}{k}=\frac{\eta g_s\,n}{2\pi\ri}.
\ee
Positive $\nu$ will correspond to symmetric representations and negative $\nu$ to 
antisymmetric ones. In the 't Hooft limit, for fixed $N/k$, the two scalings are clearly 
equivalent.

The calculation of (\ref{weta}) is very similar to the calculation of partition functions of 
$n$ bosons or fermions in the {\it canonical} ensemble, where $n$ is fixed and large. 
But at large $n$, in the thermodynamic limit, this calculation can be done 
as well in the {\it grand canonical} ensemble. We then introduce the fugacity $z$ 
and consider the grand-canonical partition function, using the expression for the 
determinant as the generating function of the characters
\be
\label{gcpf}
\Xi_{\eta} (z)=\sum_{n\ge 0} z^n W_n^{\eta}
=\Big\langle \det\left(1-\eta z\,U\right)^{-\eta} \Big\rangle
=\left\langle \exp\Bigg( \sum_{\ell \ge 1} {\tr \, U^{\ell} \over \ell} \eta^{\ell-1}z^{\ell} \Bigg) \right\rangle. 
\ee
The average value of $n$ in this ensemble is given by (we remove the average 
notation here, as is standard in the grand canonical formalism)
\be
n=z{\partial \over \partial z} \log \Xi_\eta. 
\label{saddle}
\ee
This is inverted to determine the fugacity as a function of the number of particles
\be
z_*=z_*(n),
\ee
and then the original VEV can be calculated, in a saddle point approximation, as
\be
W_n^{\eta}\approx  z_*^{-n} \Xi_\eta(z_*)=\left\langle \exp\Bigg(-n \log z_*+ \sum_{\ell \ge 1} {\tr \, U^{\ell} \over \ell} \eta^{\ell-1}z_*^{\ell} \Bigg) \right\rangle.
\ee

For convenience, let us henceforth absorb $Y=\eta z$. 
It can be seen that, at leading order in large $N$, the grand-canonical partition 
function (\ref{gcpf}) is given by disconnected planar graphs. Therefore
\be
\label{defgY}
\Xi_\eta(Y) \approx \exp\left(\frac{\eta}{g_s}\,g(Y)\right),
\qquad
g(Y)= g_s\sum_{\ell \ge 1} {\langle \tr \, U^{\ell} \rangle_0 \over \ell} \,Y^{\ell},
\ee
where the subscript $0$ refers to the planar part. We now observe that the function 
$g(Y)$ is related to the planar resolvent in the lens space matrix model (\ref{resolv}) 
and (\ref{explicitRes}) by
\be
\ba
Y\,\frac{\p}{\p Y}g(Y)
=&\frac{1}{2}\left(\omega_0(Y^{-1})-t\right)\\
=&-\log \left( \frac{1}{2}\left[ {\sqrt{(Y+b )(Y+1/b)}}+{\sqrt{(Y-a )(Y-1 /a)}}\right] \right).
\ea
\label{gderiv}
\ee
Note that compared to $\omega_0$ in (\ref{explicitRes}), the sign between the two 
square roots is reversed. Integrating this equation we get
\be
\label{fullintegral}
g(Y)
=-\int_0^{Y} \frac{\rd Y'}{Y'}
\log \left( \frac{1}{2}\left[ {\sqrt{(Y'+b )(Y'+1/b)}}+{\sqrt{(Y'-a )(Y'-1 /a)}}\right] \right).
\ee
The initial point of integration is chosen to be $Y=0$, since around that point the 
integrand approaches a constant $\zeta/2+\CO(Y)$. This guaranties that for small 
$Y$ the result of the integration will be proportional to the $1/2$ BPS Wilson loop 
(\ref{12planar}).

The saddle point equation (\ref{saddle}) determining the mean value of $n$ is then given by 
\be
\nu =\frac{1}{2\pi\ri}\, Y{\partial \over \partial Y}  g(Y). 
\ee
{\em i.e.}, (\ref{gderiv})
\be
\re^{-2\pi\ri \nu}
=\frac{1}{2}\left[\sqrt{(Y_* +b)(Y_* +1/b)}+\sqrt{(Y_* -a)(Y_* -1/a)}\right],
\qquad
Y_*=\eta z_*.
\ee
This can be solved explicitly in terms of $\beta$, $\zeta$ or alternatively in terms of 
$B$ and $\kappa$. The solution reads
\be
\label{exsol}
Y_*=\frac{\ri\kappa \,\re^{- \pi\ri(2\nu+B)}}{4\sin(2 \pi(\nu+B))}
\left(1-\sqrt{1-\frac{16\sin(2\pi\nu) \sin(2 \pi(\nu+B))}{\kappa^2}}\right).
\ee
The choice of sign is such that $Y_*=0$ when $\nu=0$. 
We will write
\be
W_n^{\eta} \approx \exp\left(A_\eta/g_s\right)
\ee
where $A_\eta$, which is identified with the action of a brane probe in the large 
$N$ string/M-theory dual, is given by 
\be
\label{action}
A_\eta =-2\pi\ri\eta\nu \log(\eta Y_*) +\eta g(Y_*).
\ee

In the original variables, in terms of $\omega_0$, the integral (\ref{fullintegral})
is from infinity to a finite position $Y_*^{-1}$, and represents 
the effect of adding a single eigenvalue to the system. This fits with the standard 
dictionary \cite{llm} identifying a brane with a single eigenvalue.

This integral gives an expression for the action of the giant Wilson loop, in the limit 
(\ref{giantl}) which is exact as a function of the 't~Hooft couplings. 
The derivatives of this integral with respect to $\beta$ and $\zeta$ can be evaluated in 
closed form, as in (\ref{derperzeta}), in terms of {\em incomplete} elliptic integrals. 
The resulting expression can then be studied at the different limits of the ABJM theory 
as done for other observables in earlier sections.

If we go to the conifold limit, setting $\lambda_2=0$, we get an expression 
for the giant Wilson loop in Chern--Simons theory on $\IS^3$. In that case there exists 
an exact expression for the Wilson loop for all $n$. As we show in 
Appendix~\ref{sec:cs-giant}, the above derivation in this limit indeed 
reproduces the CS answer.

We will now discuss the expansion of the result for the giant Wilson loop for large $\kappa$, 
since this is the strong coupling limit in which one makes contact with the AdS geometry 
\cite{df}. In terms of $B$ and $\kappa$, the integral (\ref{fullintegral}) reads
\be
\label{zetain}
\ba
g(Y_*)=-\int_0^{Y_*} {\rd Y'\over Y'} 
\log\bigg(\frac{1}{2}\bigg[&\sqrt{(1+Y' )^2-\re^{\pi\ri B} Y'(\kappa-4\ri\sin(\pi B))}
\\&
+\sqrt{(1-Y')^2 -\re^{\pi\ri B} Y'(\kappa+4\ri\sin(\pi B))}\bigg]\bigg)
\ea
\ee
where $Y_*$ is given in (\ref{exsol}). 

Expanding $Y_*$ at leading order at large $\kappa$ we get
\be
Y_*=2\ri\,\re^{-\pi\ri(2\nu+B)}\frac{\sin(2\pi\nu)}{\kappa}+\CO(\kappa^{-2}) 
=\frac{1-\re^{-4\pi\ri\nu}}{\kappa}\,\re^{-\pi\ri B}+\CO(\kappa^{-2}) 
\ee
This suggests rescaling $Y$ in the integral (\ref{zetain}) by $\kappa$, which allows for a 
systematic expansion in powers of $\kappa^{-1}$. At leading order the integral becomes
\be
g(Y_*)=-\int_0^{Y_*} {\rd Y'\over Y'} 
\left(\log\sqrt{1-\re^{\pi\ri B} \kappa Y'}+\CO(\kappa^{-1})\right).
\ee
This yields
\be
g(Y_*)=\frac{1}{2}\text{Li}_2\!\left(\re^{\pi\ri B} \kappa Y_*\right)+\CO(\kappa^{-2})
=\frac{1}{2}\text{Li}_2\!\left(1-\re^{-4\pi\ri\nu}\right)+\CO(\kappa^{-2})
\ee
Another way to get this estimate is to notice that the highest powers of $\zeta$ in the series expansion in $y$ of $g(y)$ are captured by
\be 
g(y)={1\over 2} \text{Li}_2(\zeta y) +\cdots. 
\ee
Using the dilogarithem identity (\ref{dilog1}) 
we conclude that the action (\ref{action}), written in terms of the original variable $n$, is
\be
\label{fresult}
\frac{1}{g_s}A_\eta
=n \pi {\sqrt{2\hat\lambda}} +\frac{n\pi \ri}{2}(2B-1+\eta) 
+ {\eta k \over 4\pi\ri} \left( {\pi^2\over 6} 
-\text{Li}_2\!\left(\re^{-4\pi\ri n/k}\right)\right) +{\CO}(\hat\lambda^{-1/2}, \re^{-2\pi {\sqrt{2 \hat\lambda}}}).
\ee
Notice that this formula does not display the exchange symmetry $n \leftrightarrow N-n$ for the antisymmetric case $\eta=-1$. This is because this symmetry is not present 
for the antisymmetric super-representation, as pointed out in \cite{bars}. 

The leading order in $\lambda$ in (\ref{fresult}) is as expected, {\em i.e.}, 
$n$ times the action of the fundamental string (and $n$ times an extra framing 
factor). The non-trivial dependence on $\nu$ only appears at subleading order 
in $\lambda$, and therefore will not be visible in the supergravity approximation. 
As mentioned above, there are no known $1/2$ BPS brane solutions carrying 
less than $k/2$ units of electric charge other than fundamental strings. So we 
expect that the above action describes the interaction of these coincident strings.

For $n$ a multiple of $k/2$ (or of $k$, if it is odd), we see from (\ref{exsol}) that 
$Y_*=0$ and the integral (\ref{fullintegral}) is over a full cycle. The argument of the 
dilogarithm in (\ref{fresult}) is unity, canceling the $\pi^2/6$ term. Since $Y^*$ 
passed through one of the cuts $\CC_1$ or $\CC_2$, it is now on a different 
sheet, and exactly at the branch point of the logarithm in $\omega_0(Y^{-1})$. 
This happens exactly for the value of $n$ where the strings describing the Wilson 
loop can be replaced by D2-branes, which are the string theory incarnation of 
the vortex loop operators \cite{vortex}. This suggests that the vortex loop operators 
are related to eigenvalues along the logarithmic branch-cut. It is possible to 
use our formalism to calculate the perturbative and instanton corrections to the 
these configurations and it would be interesting to understand further their significance 
in the matrix model.

\section*{Acknowledgements}
We would like to thank Ofer Aharony, Massimo Bianchi, Andrea Brini, Aristos Donos, 
Valentina Forini, Sean Hartnoll, Aki Hashimoto, 
Diego Hofman, Anton Kapustin, Albrecht Klemm, Joe Minahan, Juan Maldacena, 
Vasily Pestun, Jan Plefka, Olof Sax and Christoph Sieg
for stimulating discussions. N.D. would like to thank Nordita and the 
Erwin Schr\"odinger International Institute for their 
hospitality in the final stages of this project. M.M. would like to thank the Physics Department at Harvard University and 
the Erwin Schr\"odinger International Institute for hospitality. The work of M.M. and of P.P. is supported by the Fonds National Suisse.

\appendix

\sectiono{Normalization of the ABJM matrix model}
\label{sec:normalize}

Here we shall fix the overall normalization of the matrix model. As explained in the beginning of 
Section~\ref{sec:MM}, to fix the normalization we must fix the coefficient of the $\cosh$ in the denominator. This term appears as a consequence of 
integrating out the matter hypermultiplets at one-loop. For general supersymmetric Chern--Simons-matter theories, the contribution 
of a hypermultiplet in representation $R$ is given by \cite{kapustin}
\begin{equation}
\label{za}
\log \,  Z [a]=\log \prod_\rho \prod_{n=1}^\infty\left(\frac{n+1/2+\ri\rho(a)}{n-1/2-\ri\rho(a)}\right)^n 
\end{equation}
where $\rho$ are the weights of the representation, and $a$ is the element in the Cartan algebra given by
\be
a={1\over 2\pi}\text{diag}\left(\mu_1, \cdots, \mu_{N_1}, \nu_1, \cdots, \nu_{N_2} \right).
\ee
In \cite{kapustin} the one-loop determinant is evaluated up to a multiplicative constant, 
\be
Z [a]=\prod_{\rho} \left ( C \cosh\left( \pi \rho(a) \right) \right) ^{-1/2}.
\ee
The constant $C$ can be determined by setting $a=0$ in (\ref{za})
\be
\label{logc}
-{1\over 2} \log C=\log\prod_{n=1}^\infty\left(\frac{n+1/2}{n-1/2}\right)^n.
\ee
This is a divergent constant, but as usual when considering determinants on compact manifolds, we can 
compute it by using $\zeta$-function regularization. Let us define
\be
\zeta_Z (s) =\sum_{n=1}^{\infty} \left( {n \over \left( n+{1\over 2}\right)^s} -{n \over \left( n-{1\over 2}\right)^s} \right). 
\ee
The regularization of the quantity appearing in (\ref{logc}) is then $-\zeta'_Z(0)$. An elementary calculation shows that
\be
\zeta_Z(s) = -\left(2^s-1 \right) \zeta(s)
\ee
where $\zeta(s)$ is the standard Riemann zeta function. Therefore, 
\be
-\zeta'_Z(0)=-{\log 2\over 2} 
\ee
and $C=2$.

\sectiono{Giant Wilson loops in Chern--Simons theory}
\label{sec:cs-giant}

Chern--Simons theory on $\IS^3$ is a particular case of the lens space matrix 
model when $b=1$ and the second cut collapses to zero size, {\em i.e.}, 
$t_1=t$, $t_2=0$. It gives the leading behavior of the Wilson loop in ABJM theory 
when $\lambda_2\ll\lambda_1$, as discussed in Section~\ref{sec:conifold}.

Here we consider the behavior of the giant Wilson loops, those in high 
dimensional symmetric or antisymmetric representations presented in 
Section~\ref{sec:giant}, in this limit.
In this case it is easy to calculate explicitly the 
action (\ref{action}), since the integral 
\be
\label{gyS3}
g(Y)=-\int_0^{Y} {\rd Y'\over Y'}\log(h(Y'))\,,
\qquad
h(Y)=\frac{1}{2}\left[1+Y+\sqrt{(1+Y)^2 -4\re^t Y}\right]
\ee
can be obtained in closed form
\be
\ba
\label{gS3ex}
g(Y)=&\,{\pi^2\over 6} -{1\over 2} \log^2(h(Y))
+\log (h(Y))\Bigl( \log\big(1-\re^{-t}h(Y)\big)-\log(1-h(Y))\Bigr)\\
&-\text{Li}_2(h(Y)) +\text{Li}_2\big(\re^{-t}h(Y)\big) -\text{Li}_2 (\re^{-t}). 
\ea
\ee
Here we used the dilogarithm identity
\be
\label{dilog1}
\text{Li}_2(1-x)={\pi^2 \over 6} -\text{Li}_2(x) -\log (x) \log (1-x).
\ee
The solution of the saddle point equation (\ref{saddle}) is obtained by setting in (\ref{exsol})
\be
\kappa=-4\ri\sinh\frac{t}{2}\,,\qquad
B=\frac{t}{2\pi\ri}+\frac{1}{2}
\ee
and we find   
\be
\label{CSz}
Y_*=-\frac{1-\re^{-2\pi\ri\nu }}{1-\re^{2\pi\ri\nu+t}}\,.
\ee
The action (\ref{action}) is
\be
\label{exactA}
\ba
\eta\,A_\eta&=-2\pi\ri\nu \log(\eta Y_*) + g(Y_*)
\\&=-2\pi\ri\nu \log \eta -2\pi^2 \nu^2 +2\pi\ri\nu t
+{\pi^2 \over 6}+\text{Li}_2\big(\re^{2\pi\ri\nu-t}\big)
-\text{Li}_2\big(\re^{2\pi\ri\nu}\big) -\text{Li}_2\big(\re^{-t}\big).
\ea
\ee
Notice that this expression is exact in $t$. 

We can test (\ref{exactA}) in all details against a direct calculation of correlators. Indeed, the VEVs $\langle \tr_{R}\,U \rangle$ for the 
Chern--Simons matrix model on $\IS^3$ are proportional to 
quantum dimensions (see for example \cite{leshouches}):
\be
\label{CS-WL}
\langle \tr_{R}\,U \rangle =q^{\kappa_R/2 +\ell(R) N/2} \text{dim}_q (R).
\ee
In this equation, 
\be
q=\re^{g_s},
\ee
$\ell(R)$ is the number of boxes in $R$, and $\kappa_R$ is the framing factor, given by
\be
\kappa_R=\sum_i l_i(l_i-2i+1),
\ee
where $l_i$ are the lenghts of the rows in the diagrams. The quantum dimensions of the symmetric and antisymmetric representations are given by
\be
\label{sas}
\text{dim}_q(\CR_n^\eta)={q^{\eta n(n-1)/ 4}\re^{nt/2} \over [n]!} \prod_{i=1}^n (1-\re^{-t} q^{-\eta(i-1)}),
\ee
where
\be
[n]!=\prod_{i=1}^n (q^{i/2}-q^{-i/2})=q^{{1\over 4}n(n+1)} \prod_{i=1}^n (1-q^{-i}). 
\ee
At large $n$ we rescale
\be
\xi={i\over n}, \qquad q^{-i} =\exp( -g_s i) \rightarrow \re^{-2\pi\ri\eta\nu \xi}
\ee
so that  
\be
\log([n]!) \approx {1\over g_s} \left(-\pi^2 \nu^2 
+2\pi\ri\eta \nu \int_0^1 \rd \xi \log(1-\re^{-2\pi\ri\eta\nu \xi}) \right).
\ee
This gives the following contribution to the action
\be
\label{factorial}
\pi^2\nu^2 +{\pi^2\over 6} -\text{Li}_2\big(\re^{-2\pi\ri\eta\nu}\big)
=\eta\left(\pi^2\nu^2-2\pi\ri\nu\log\eta+\frac{\pi^2}{6}-\text{Li}_2\big(\re^{-2\pi\ri\nu}\big)\right). 
\ee
To derive the expression on the right hand side we used, for $\eta=-1$ the dilogarithm identity 
\be
\text{Li}_2(\re^{x}) =-\text{Li}_2(\re^{-x}) +{\pi^2 \over 3} -{x^2 \over 2}\pm\pi\ri x.
\ee

The product in the numerator of both the symmetric and antisymmetric representations 
can be written in a unified form as
\be
\label{q-dim}
2\pi\ri\eta\nu  \int_0^1 \rd \xi \log(1-\re^{-t} \re^{-2\pi\ri\nu \xi})  
=\eta\left(\text{Li}_2(\re^{-t-2\pi\ri\nu}) -\text{Li}_2(\re^{-t})\right). 
\ee
The prefactors in (\ref{CS-WL}) and (\ref{sas}) contribute
\be
\eta(-3\pi^2\nu^2+2\pi\ri\nu t).
\ee
Together with (\ref{factorial}) and (\ref{q-dim}) this exactly reproduces (\ref{exactA}).

In the antisymmetric representation the result can also be written as
\be
\label{antieasy}
-2\pi\ri\nu (t+2\pi\ri\nu) +{\pi^2\over 6}
+\text{Li}_2(\re^{-t})-\text{Li}_2(\re^{-t-2\pi\ri\nu}) -\text{Li}_2(\re^{2\pi\ri\nu}).
\ee
This expression agrees at leading order with the D6-brane calculation 
(\ref{16giant}) and should be the full answer in the limit of $\lambda_2=0$. 
In this expression we see the expected symmetry \cite{yamaguchi}
\be
n \leftrightarrow N-n
\ee
which is 
\be
2\pi\ri\nu \leftrightarrow -t-2\pi\ri\nu.
\ee

\end{document}